XeLatex=\RequirePackage{lineno}
\documentclass[aps, prl,
floatfix,twocolumn,superscriptaddress,
]{revtex4-2}
\usepackage{t1enc}
\usepackage[T1]{fontenc}
\usepackage{lmodern}
\usepackage{graphicx}
\usepackage{epsfig}
\usepackage{amssymb}
\usepackage{amsmath}
\usepackage{dcolumn}
\usepackage{bm}
\usepackage{color}
\usepackage{float}
\usepackage{hyperref}
\usepackage{natbib}
\usepackage{footnote}
\usepackage{ulem}
\usepackage{hhline}
\usepackage[export]{adjustbox}
\begin{document}

\title{Propagating charge carrier plasmon in Sr$_2$RuO$_4$}
\author{Martin\,Knupfer}
\affiliation{Leibniz Institute for Solid State and Materials Research  Dresden, Helmholtzstr. 20, D-01069 Dresden, Germany}
\author{Fabian\,Jerzembeck}
\affiliation{Max Planck Institute for Chemical Physics of Solids, N\"othnitzer Str. 40, 01187 Dresden, Germany}
\author{Naoki \,Kikugawa}
\affiliation{National Institute for Materials Science, Tsukuba 305-0003, Japan}
\author{Friedrich\,Roth}
\affiliation{Institute of Experimental Physics, TU Bergakademie Freiberg, Leipziger Str. 23, 09599 Freiberg, Germany}
\affiliation{Center for Efficient High Temperature Processes and Materials Conversion,
TU Bergakademie Freiberg, 09599 Freiberg, Germany}
\author{J\"org\,Fink}
\affiliation{Leibniz Institute for Solid State and Materials Research  Dresden, Helmholtzstr. 20, D-01069 Dresden, Germany}
\affiliation {Institut f\"ur Festk\"orperphysik,  Technische Universit\"at Dresden, D-01062 Dresden, Germany}

\date{\today}

\begin{abstract}\

We report on studies of charge carrier plasmon excitations in Sr$_2$RuO$_4$ by transmission Electron Energy-Loss Spectroscopy. In particular, we present results  on the plasmon  
dispersion and its width as a function of momentum transfer. The dispersion can be qualitatively explained in the framework of RPA calculations, using an unrenormalized tight-binding band structure.  The constant long-wavelength width of the plasmon  indicates, that it is caused by a decay into inter-band transition and not by quantum critical fluctuations. The results from these studies on a prototypical ``bad`` metal system show that the long-wavelength plasmon excitations near 1 eV are caused by resilient quasiparticles and are  not influenced by correlation effects.
\end{abstract}


\maketitle

\paragraph{Introduction.} 
Unconventional superconductors are one of the most active research fields in solid state physics. The reason for this is not only the observed high superconducting transition temperature. Very often, these materials are termed ``bad`` metals because of their  strange normal state properties. The latter are caused by strong electron-electron interactions leading to correlation effects  connected with quantum critical behavior. The renormalizations of single-particle excitations are described by the complex self-energy $\Sigma$, the real and the imaginary part of which are related to the mass renormalization $m^*/m$ and the scattering rate $\Gamma$, respectively.
\par
One of the most studied ``bad`` metals are the doped cuprates. There are numerous studies of single-hole excitations in these materials by ARPES providing detailed knowledge of the self-energy yielding a  mass renormalization of the conduction band of about two~\cite{Damascelli2003}. Furthermore, there are studies of the spin excitations compiled  in Ref.~\cite{Plakida2016}. On the other hand, there are only a few studies on momentum dependent charge degrees of freedom. There is an ongoing discussion, how the renormalization of the single-particle excitations, i.e., the mass renormalization and the finite scattering rate, influences the existence, the dispersion, and the lifetime  broadening of  two-particle charge excitations (plasmons) in the cuprates. 
\par
Several theoretical studies~\cite{Tohyama1995,Khaliullin1996,Eder1995} have predicted, that  spin excitations strongly influence the charge excitations. On the other hand, the observation of a continuum instead of the high-energy plasmon was discussed for quantum critical systems based on
holographic theories~\cite{Romero-Bermudez2019}, which predicted, different from the classical Landau damping~\cite{Mahan2000}, a strong ``over-damping`` even for the long-wavelength plasmons. These theoretical investigations seem to be supported by Electron Energy-Loss Spectroscopy (EELS) in reflection (R-EELS) studies of cuprates~\cite{Vig2017,Mitrano2018,Husain2020}  in which not well-defined propagating plasmons or even only a continuum of excitations were detected. On the other hand, there are EELS studies in transmission (T-EELS)~\cite{Raether1965,Daniels1970,Schnatterly1979,Schattschneider1986,Fink1989,Roth2014} on cuprates~\cite{Nuecker1989,Nuecker1991,Vielsack1990,Grigoryan1999,Paasch1999,Roth2020} where the dispersion of well defined plasmons could be explained in terms of collective excitations within an  {\it unrenormalized} band structure.
\par 
To better understand the problem, in the present paper, we extend the studies to Sr$_2$RuO$_4$, a ``bad`` metal with an even higher average effective mass of about four,  using T-EELS.
Sr$_2$RuO$_4$ is a prototype of  unconventional superconductors~\cite{Maeno1994,Mackenzie2003}. The superconducting order parameter  is still a matter of controversy~\cite{Sharma2020}. The strongly correlated normal state electronic structure has been studied by optical spectroscopy~\cite{Katsufuji1996,Hildebrand2001,Pucher2003,Stricker2014,Stricker2015}, by R-EELS~\cite{Husain2020}, by tunneling spectroscopy~\cite{Wang2017}, and by ARPES~\cite{Wang2004,Ingle2005,Zabolotnyy2013,Akebi2019,Tamai2019}, in some cases supported by DFT or combined DFT plus dynamical mean field theory (DFT+DMFT) calculations~\cite{Singh1995,Liebsch2000,Tamai2019}. The latter two experimental techniques probe essentially one-particle (electron or hole) properties whereas the first two methods probe two-particle excitations (hole plus electrons).
\par 
Using slices of the Sr$_2$RuO$_4$ crystal cut perpendicular to the ${\bf c}$ axis together with an orientation of the ${\bf c}$ axis parallel to the electron beam, in the present T-EELS measurements, the momentum can be varied within the (${\bf a,b}$) plane. The momentum parallel to the beam is negligible. Optical spectroscopy measuring excitations with  zero momentum transfer~\cite{Katsufuji1996,Hildebrand2001,Stricker2014,Stricker2015} found for a photon polarization perpendicular to the ${\bf c}$-axis a well pronounced plasmon near 1.5 eV with a plasmon width of $\approx$ 1 eV. For a polarization parallel to the ${\bf c}$-axis, a low energy plasmon near 0.01 eV was detected~\cite{Katsufuji1996,Hildebrand2001,Pucher2003}. Different from this result, R-EELS study detected near 1 eV a continuum and at low energy a dispersing acoustic plasmon~\cite{Husain2020}, which was ascribed to a ``Demon``  plasmon on the basis of interacting excitations of heavy and light electrons. 
\par
The existence of a high-energy and a low-energy plasmon was already predicted from a calculation of the  high in-plane and the low out-of-plane Fermi velocities on the basis  of band structure calculations~\cite{Singh_1995}. 
\par
The main result of the present study is the observation of a well-defined propagating plasmon which in the \textit{non-local} long-wavelength limit can be qualitatively explained by  RPA calculations based on an \textit{unrenormalized} DFT band structure. This indicates that the low-momentum dispersion is not influenced by the 
\textit{local} on-site Coulomb interaction. 
The results also demonstrate that single hole excitations (measured by ARPES) are much stronger screened than two-particle electron-hole excitations (measured e.g. by T-EELS). 

\paragraph{Experimental.} 

Sr$_2$RuO$_4$ crystals  were grown using the traveling floating-zone
method~\cite{Bobowski2019}. The superconducting transition temperature of the sample was $T_c=$1.5 K.
For the EELS measurements, thin films with a thickness of $\approx$\,100\,nm were cut perpendicular to the crystal ${\bf c}$-axis from these single crystals
using an ultramicrotome equipped with a diamond knife. The films were then put onto standard transmission electron microscopy grids and
transferred into the spectrometer. The measurements were carried out at a sample temperature of $T=$~20 K with a dedicated transmission electron energy-loss
spectrometer \cite{Fink1989,Roth2014} employing a primary electron energy of 172\,keV. The energy and momentum resolution was set to  $\Delta
E$\,=\,80\,meV and $\Delta q$\,=\,0.035\,\AA$^{-1}$, respectively. In all measurements, the momentum transfer was parallel to the (${\bf a,b}$) plane. Before measuring the loss-function, the thin films have been characterized by \textit{in-situ} electron diffraction, in order to orient the crystallographic axis with respect to the
transferred momentum.

\par

\paragraph{Calculations.}
For the calculation of the complex dielectric function we used the Ehrenreich-Cohen expression~\cite{Ehrenreich1959,Giuliani2005}:
\begin{equation}\label{EC}
\epsilon (\omega,{\bf q})=\epsilon _{\infty}-\frac{1}{q^2}\chi_0(\omega,{\bf q})
\end{equation}
with 
\begin{equation}\label{CHI}
\chi_0(\omega,{\bf q})\propto \frac{1}{q^2}M^2\int_{-\infty}^{\infty} \frac{F_k \Delta E}{\omega^2-\Delta E^2+i\Gamma_P \omega } d{\bf k}.
\end{equation}
Here $\Delta E=E_{\bf k+q}-E_{\bf k}$  where $E_{\bf k}$ describes the band structure, $F_{\bf k}$ is the Fermi function, $\Gamma_P$ is the lifetime broadening,  and M is the matrix element. Using the complex dielectric function we then calculate the loss function  $\Im[-1/\epsilon(\omega ,{\bf q})]$. The maximum of the loss function as a function of momentum {\bf q}  which occurs at  
$\Re[\epsilon(\omega,{\bf q})]=0$
 yields  the plasmon dispersion. In the calculation we use a constant matrix element, the absolute value being determined by the condition that the plasmon energy for $q=0$  agrees with that derived from optical spectroscopy. The small momentum dependence of the matrix element is supported by the observation that for  momentum $ q\leq $0.5 \AA$^{-1}$   the longitudinal f-sum rule~\cite{Mahan2000} is reduced by less than $20 ~\%$. Similar calculations were performed for poly-acetylene~\cite{Neumann1987}, La$_2$CuO$_4$~\cite{Vielsack1990}, optimally doped  Bi$_2$Sr$_2$CaCu$_2$O$_8$~\cite{Nuecker1991,Grigoryan1999}, and for the ladder compound Ca$_{x}$Sr$_{14-x}$Cu$_{24}$O$_{41}$~\cite{Roth2020}. 
Interestingly, in all cases the low-energy dispersion could be well described using a simple {\it unrenormalized} tight-binding band structure. 

\par

In Sr$_2$RuO$_4$ the conduction band is composed of 3 Ru 4$d$ $t_{2g}$ bands, the narrow quasi-1D  $\alpha$ and $\beta$ bands, and the wider quasi-2D $\gamma$ band, having, without taking into account spin-orbit interaction, predominantly $xz$, $yz$, and $xy$ orbital character, respectively~\cite{Ingle2005,Tamai2019}. For the calculation of the plasmon dispersion we used an unrenormalized tight-binding band structure~\cite{Liebsch2000}. We obtain a minimum of the difference between calculated and  experimental plasmon dispersion by a shift of 0.2 eV of the band structure to lower binding energies. For calculations using Eq.~\ref{EC} the neglect of the spin-orbit coupling is reasonable because the spin-orbit coupling is considerably smaller than the plasmon energy~\cite{Pchelkina2007,Tamai2019}.  For the calculations we used the  background dielectric function $\epsilon _{\infty}=$2.3 from optical spectroscopy~\cite{Stricker2014,Stricker2015}.
\par
Using RPA for the homogenous electron gas, the free electron long-wavelength plasmon dispersion coefficient $\alpha$ (see Eq.~3) is proportional to $ v_{42} \propto <v_F^4>/<v_F^2> \approx <v_F^2>$~\cite{Grigoryan1999,Giuliani2005}. To control  the anisotropy and  orbital character of the calculated dispersions we have calculated 
 $v_{42}$ for the  bands for ${\bf q} $ parallel to the [110] and [100] direction. 
 
\par

\paragraph{Results.}

\begin{figure}[tb]
\centering
\includegraphics[width=0.5\textwidth]{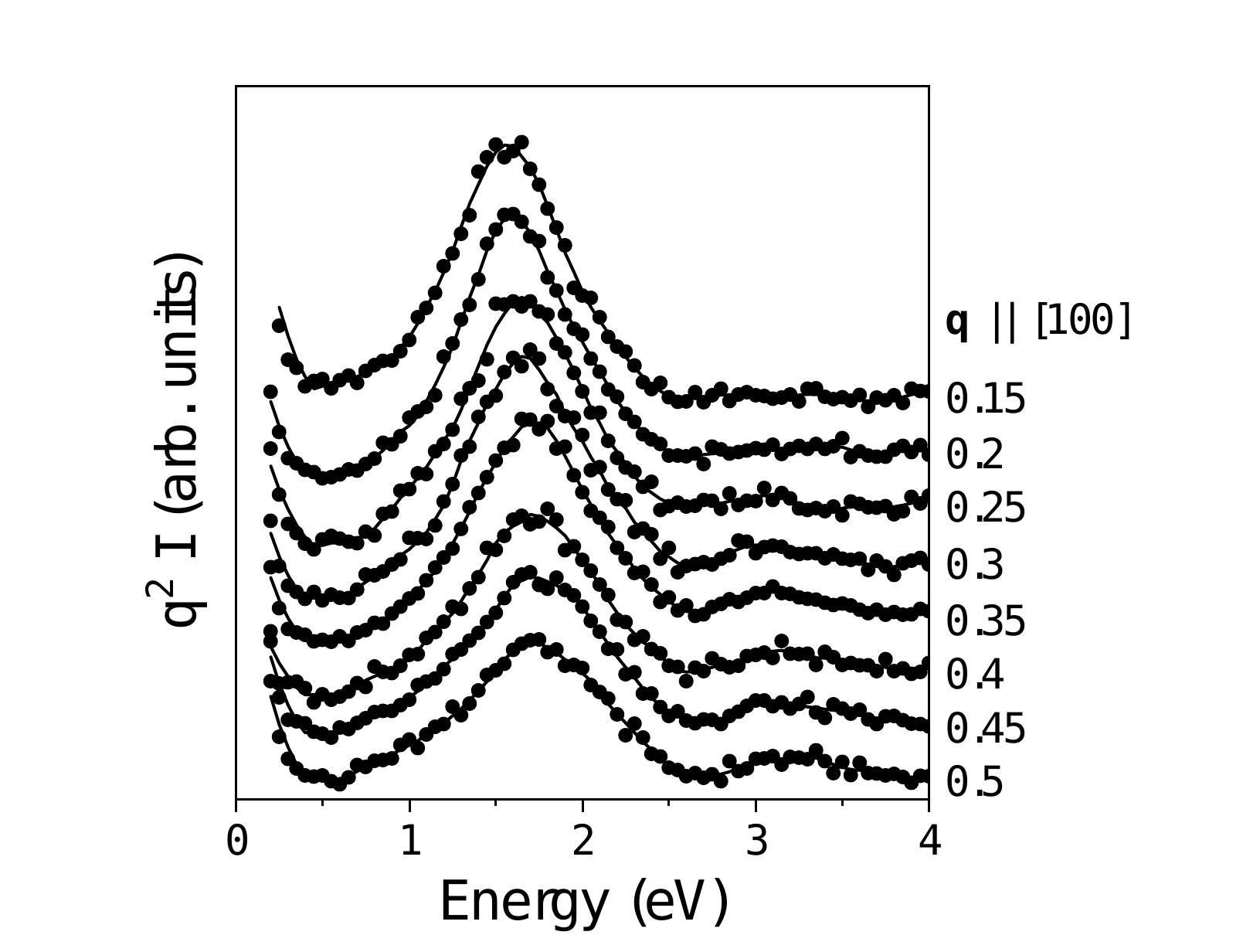}
\includegraphics[width=0.5\textwidth]{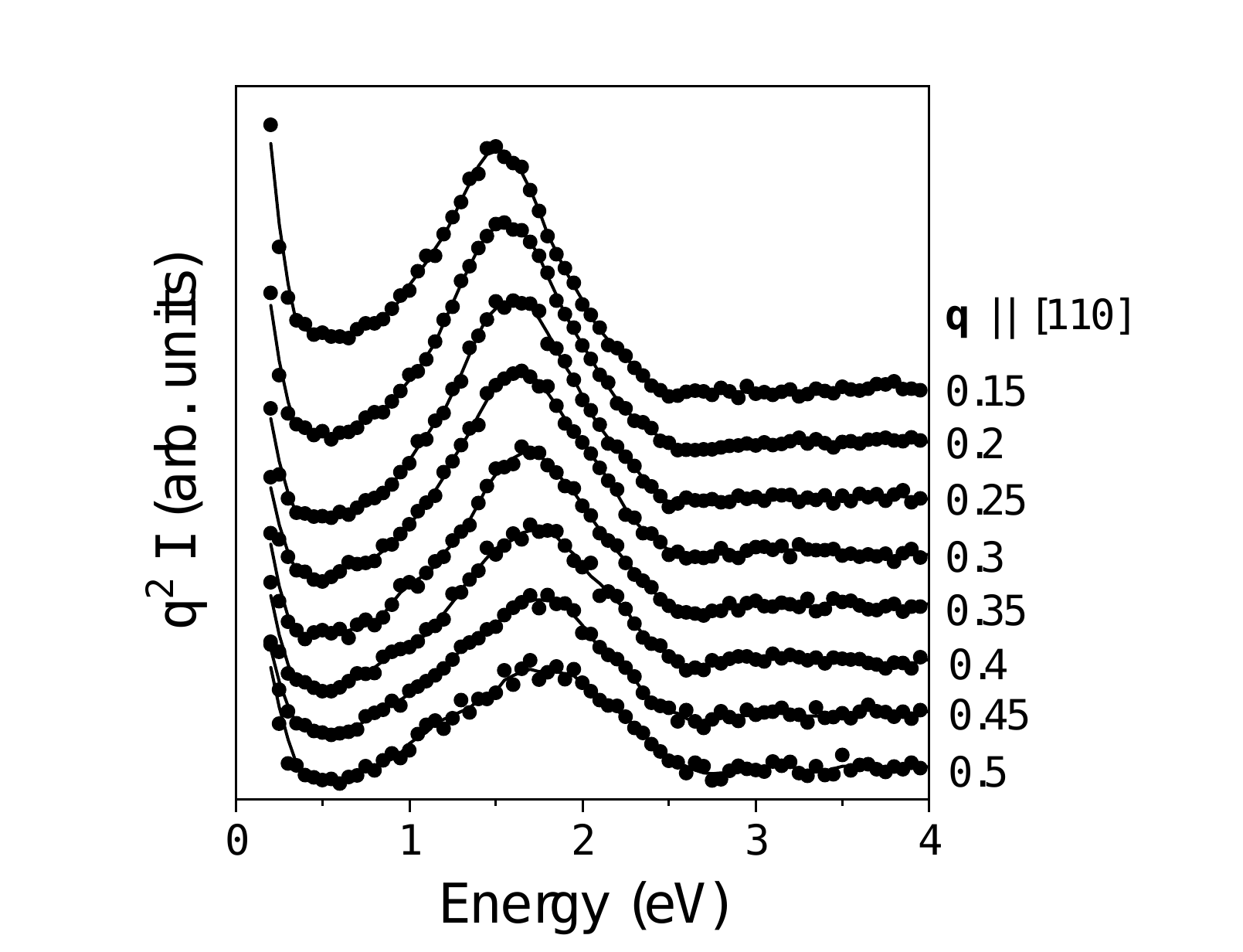}
\caption{\label{fig1}
Electron energy-loss spectra multiplied by $q^2$ ($q^2I$) of  Sr$_2$RuO$_4$ in the energy range between 0 to 4 eV for momentum transfer ${\bf q}$ between 0.15  and 0.5  \AA$^{-1}$. Upper panel: ${\bf q}$ parallel to [100] direction, lower panel: ${\bf q}$ parallel to [110] direction. Solid lines: guide to the eyes.  
}
\centering
\end{figure}
In Fig.\,\ref{fig1} we present the experimental loss spectra of  Sr$_2$RuO$_4$ for momentum transfer parallel to the [110] and the [100] direction. The strongest  excitations occur between 1.5  and 1.8 eV, which are ascribed to  plasmons of the charge carriers. At low energies, the  intensity is dominated by the quasi-elastic peak. At higher energy near 3.3 eV, a small peak possibly due to interband transitions is visible at higher momentum transfer.
\par
We derive a positive plasmon dispersion by an evaluation of the maxima of the loss data. In the long-wavelength limit the experimental and the calculated dispersion can be well described by
\begin{equation}\label{DIS}
\omega(q)=\omega_p+\frac{\hbar^2}{m} \alpha q^2. 
\end{equation}
$\omega_p$ is equal to 1.49 eV which roughly  agrees with the in-layer plasmon energy $\omega_p =$ 1.57 eV derived from optical spectroscopy~\cite{Katsufuji1996,Stricker2014,Stricker2015}. 

\begin{figure}[tb]
\centering
\includegraphics[width=0.4\textwidth]{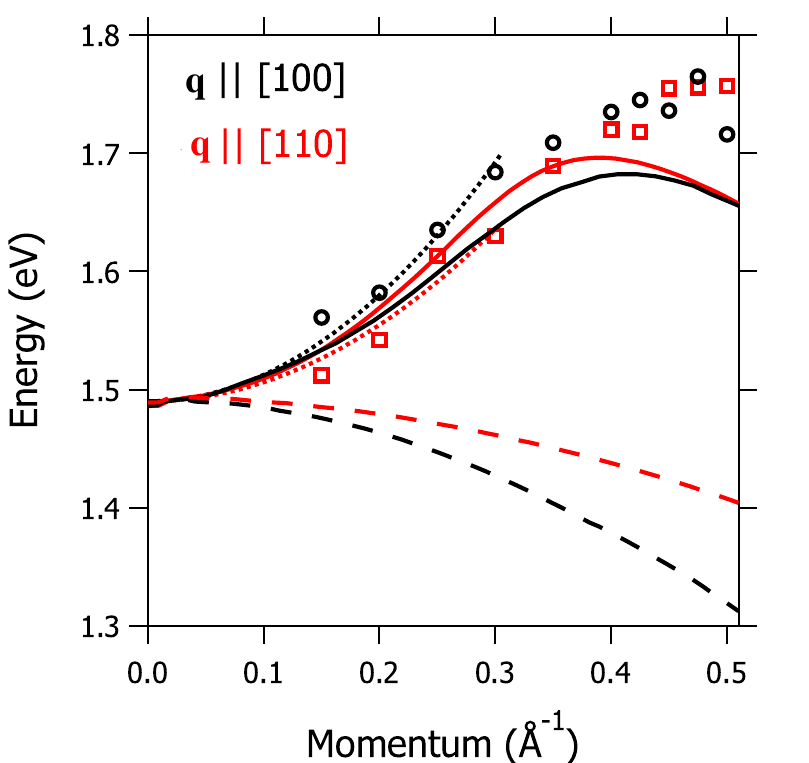}
\caption{\label{fig2}
(Color online) Experimental data of the plasmon dispersion. Circles: data for ${\bf q}$ parallel to [100]. Squares: data for ${\bf q}$ parallel to  [110]. Dotted line: least squares fit to the experimental data. Thick solid lines: results from the calculations using an effective mass of one. All black data are for ${\bf q}$ parallel to [100]. The grey (red) data are for ${\bf q}$ parallel to [110]. Dashed lines: calculations using an effective mass of four.
}
\centering
\end{figure}

The dispersions of the plasmon along the two directions together with a fit for $q \le $ 0.3 \AA$^{-1}$ is depicted in Fig.\,\ref{fig2}. 
The derived dispersion coefficients are $\alpha^{exp}_{110}=$0.22 and $\alpha^{exp}_{100}=$0.30. 

\par

In this figure we have also added the calculated plasmon dispersion along  the [110] and the [100] direction. The calculated dispersion coefficients for $q \le $ 0.3 \AA$^{-1}$ are $\alpha^{cal}_{110}=$0.25 and $\alpha^{cal}_{100}=$0.21.

\par

\begin{figure}[tb]

\includegraphics[width=0.4\textwidth]{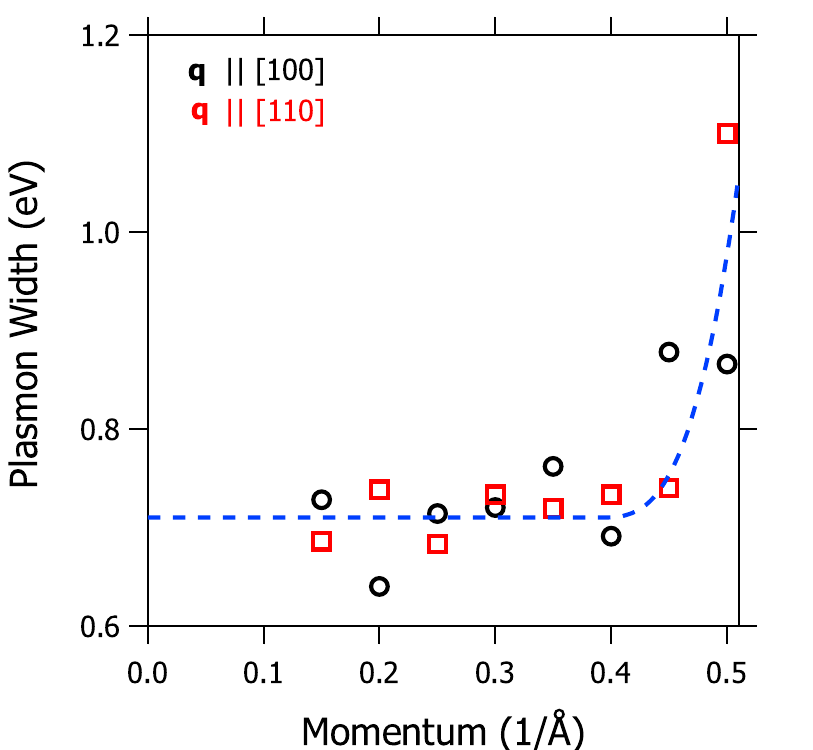}
\caption{\label{fig3}
(Color online) Experimental data of the plasmon width as a function of momentum. Black circles: data for ${\bf q}$ parallel to the [100] direction. Grey (red) squares: data for ${\bf q}$ parallel to the [110] direction. Dashed line: guide to the eyes.
}
\end{figure} 
In Fig.\,\ref{fig3} we present the plasmon width (FWHM), derived from a fit of a Gaussian plus background to the loss data shown in Fig.\,\ref{fig1}. Along the two directions, within error bars the width is constant and equal to 0.7 eV up to $q=$ 0.4 \AA$^{-1}$. Above 
$q \approx $ 0.4 \AA$^{-1}$ an increase of the plasmon width is detected.

\par
\paragraph{Discussion.} We have interpreted the  principle excitation of the loss spectra near 1.5 eV by a charge carrier plasmon. Its finite width of 0.7 eV is smaller than the plasmon energy of 1.5 eV, indicating a quasi-particle excitation. The finite dispersion signals a propagating plasmon.  We emphasize   that the width is almost constant in the studied low momentum range. This means that it is not related to electron-electron interaction  which would lead to a quadratic increase in momentum transfer~\cite{DuBois1969}. The fact that the width is smaller than the energy signals that the excitation is not overdamped due to  fluctuations in a quantum critical system, in contrast to theoretical predictions~\cite{Romero-Bermudez2019}.
Rather, it is caused, as in most metallic systems, studied by T-EELS, by a decay into interband transitions~\cite{Paasch1970,Gibbons1976,Felde1989}. These interbands are caused by a back-folding of bands from the second to the first Brillouin zone by a finite pseudo-potential.

\par
 
The observation of  a continuum near 1.5 eV and an acoustic plasmon at low energies in a recent R-EELS study~\cite{Husain2020} can be possibly explained by a large momentum transfer perpendicular to the layers leading to out-of phase collective excitations in the layer system. This view is supported by numerous theoretical studies~\cite{Grecu1973,Fetter1974,Giuliani1983,Kresin1988,Markiewicz2007,Greco2016} and by several recent Resonant Inelastic X-ray Scattering (RIXS) experiments on  layered  n-type~\cite{Ishii2017,Hepting2018,Lin2020} and p-type doped cuprates~\cite{Nag2020,Singh2022}.  In these experiments acoustic plasmons with an out-of-plane dispersion  have been detected in the low-($\omega,q$) range with a very small width of $\approx 0.1$ eV. The interpretation of the acoustic plasmons in terms of an out-of-phase excitation in layered compounds is at variance with the assignment to  a ``Demon``  plasmon~\cite{Husain2020}.
\par
Looking at  Fig.\,\ref{fig2}, the calculated plasmon dispersions, using an effective mass of one, are  close to the experimental ones. On the other hand, the small experimental anisotropy is opposite to the calculated one. The difference between the experimental and the calculated dispersion may be explained by the distortion of the bands by spin-orbit interaction~\cite{Tamai2019}, not taken into account in our RPA calculations.  An analysis of average Fermi velocities yields some information on the contribution  
of the 3  conduction bands to the plasmon dispersion: along ${\bf q}$ || to the [110] directions, the dispersion is predominantly determined by the 2D $\gamma$ band and to a lesser extent by the 1D $\alpha$ and $\beta$ bands. Along the [100] direction, both the $\gamma$ band and the $\alpha$ band contribute to the momentum dependence.  
\par
We emphasize, that we obtain a qualitative agreement between the experimental and the calculated dispersion, although for this highly correlated system, we have used in the calculation an {\it unrenormalized} tight-binding band structure. The situation is  similar to that in cuprates (see above) where the plasmon dispersion can be well described by RPA using an unrenormalized tight-binding band structure\cite{Nuecker1991,Grigoryan1999}. In the cuprates, where the band structure is renormalized by a factor of two, in the simplest approximation when the dispersion coefficient is proportional to the square of the Fermi velocity, one may expect a reduction of the  plasmon dispersion due to flattening of the bands by a factor of 4. However, in Sr$_2$RuO$_4$  the average mass renormalization of the charge carrier is near 4.5~\cite{Tamai2019}, which would lead in the above mentioned approximation to a reduction of the plasmon dispersion by a factor of about 20.  Actually, our RPA calculation using an effective mass $m^*/m=$ 4 yields a negative plasmon dispersion (see Fig.~2). As explained in previous calculations~\cite{Nuecker1991,Grigoryan1999}  negative plasmon dispersion may occur because the dispersion coefficient is composed out of a positive and a negative contribution.  A large reduction of the plasmon dispersion is not compatible with our  experimental results. Thus our present results are a clear proof  that correlation effects do not influence  the long-wavelength dispersion of the plasmon in systems which are correlated by the on-site Coulomb interaction in transition metal ions.  
\par
The weak influence of correlation effects on the plasmon dispersion can be rationalized by the following consideration.
The dispersion is a measure of the compressibility of the electron liquid~\cite{Mahan2000,Giuliani2005}. In the long-wavelength limit, the charge oscillations due to the plasmon excitation have a wavelength that  is much larger than the distance between the next nearest Ru atoms. On the other hand, the mass enhancement in the correlated materials is predominantly caused by the hopping between neighboring transition metal atoms which is needed to overcome the on-site Coulomb interaction.  In the long-wavelength limit of the plasmon excitations, the occupation number on the transition metal atoms is hardly changed and therefore the plasmon  dispersion is not influenced by correlation effects. Moreover, at high energies, the mass enhancement  is expected to be reduced because the charge carriers are no more coupled to spin excitations~\cite{Tohyama1995,Khaliullin1996,Eder1995}, which have an energy of the order of 100 meV. Furthermore, one can also understand the difference between EELS and ARPES studies. In the latter, the mass of a single hole is enhanced by an Auger-like valence band electron-hole excitation, leading in the final state to a two holes in the occupied part of the band and therefore is influenced by the on-site Coulomb interaction. 
\par
Another indication for the non-sensitivity of the plasmon excitations to correlation effects is the plasmon width, being in the long-wavelength limit nearly independent of the momentum transfer (see Fig.\,\ref{fig3}). This demonstrates, that in Sr$_2$RuO$_4$ the width is not caused by a lifetime broadening of the valence band due to a finite imaginary part of the self-energy, observed by ARPES. At $q=$0  the plasmon width must be rigorously zero in an free-electron gas~\cite{Schnatterly1979}. In the case of Al a quadratic momentum dependence of the width due to intra-band particle-hole excitations was predicted by theoretical calculations~\cite{DuBois1969}, which is  much too small compared to experimental data of Al.  Later, theoretical calculations~\cite{Paasch1970,Sturm1989} and  experiments~\cite{Gibbons1977,vom_Felde_1989,Sing1999} showed, that the plasmon width is predominantly caused by a decay of the plasmon into  inter-band transitions related to back-folded bands from the second to the first Brillouin zone by a finite pseudo-potential. This could lead to a constant or a negative $q$-dependence of the plasmon width observed for K$_{0.3}$MoO${_3}$~\cite{Sing1999} or Li~\cite{Gibbons1977}, respectively. Thus the constant width at low $q$ in Sr$_2$RuO$_4$ is probably also caused by a decay into inter-band transitions and not by quantum critical fluctuations.
\par
Moreover, there is also a more quantitative explanation of the differences of the imprint of correlation effects in ARPES and EELS studies, based on dynamical mean-field calculation theory~\cite{Deng2013,Stricker2014}. These calculations  predict for energies below about 0.1 eV a Fermi liquid behavior with a mass enhancement of about four and above that energy a strongly reduced mass enhancement. This leads for the spectral function at low energies ( 0.2 < $\omega$ > -0.5 eV) to a strongly renormalized dispersion and at higher energies to an unrenormalized dispersion of resilient quasiparticles which determine the ``bad`` metal properties. The observation of a plasmon dispersion, related to high-energy excitations, which are not influenced by correlation effects, is a clear indication of such resilient quasiparticles.
\par
A comparison of the calculated susceptibility (not shown) and the plasmon dispersion shows that independent of the momentum direction, the plasmon merges into the single particle excitations near $q=$ 0.4 \AA$^{-1}$. At this momentum also an additional broadening of the experimental plasmon width is observed (see Fig.\,\ref{fig3}). This behavior is very similar to that observed in nearly-free electron metals~\cite{Felde1989}.
\par
There is a certain flattening of the experimental dispersion at higher momentum transfer. As it also appears  in the RPA calculations, we conclude that it is not caused by correlation induced local field corrections. Rather we infer that it is caused by the finite band width of the conduction bands. To obtain information on the dynamic local field correction experiments at higher momentum transfer are required.
\par
In summary, we have observed in Sr$_2$RuO$_4$ a propagating plasmon, which is at low momentum and at high energies determined by resilient quasiparticles which are not renormalized. Extending future EELS measurement to higher momentum and into the low-energy range using an improved energy resolution will provide valuable information on the renormalized part of the spectral function and on the momentum dependence of correlation effects.

\section{ ACKNOWLEDGMENTS}
We thank  Peter Abbamonte, Bernd B\"uchner, Luis Craco, Stefan-Ludwig Drechsler, Axel Lubk, and Andrew P. Mackenzie for helpful discussions, and Marco Naumann for technical assistance.
This work is supported by a KAKENHI Grants-in-Aids for Scientific Research (Grant Nos. 18K04715,  21H01033, and 22K19093), and Core-to-Core Program (No. JPJSCCA20170002) from the Japan Society for the Promotion of Science (JSPS) and by a JST-Mirai Program (Grant No. JPMJMI18A3).

\bibliographystyle{apsrev4-2}
\raggedright
\bibliography{EELS}

\begin{thebibliography}{64}%
\makeatletter
\providecommand \@ifxundefined [1]{%
 \@ifx{#1\undefined}
}%
\providecommand \@ifnum [1]{%
 \ifnum #1\expandafter \@firstoftwo
 \else \expandafter \@secondoftwo
 \fi
}%
\providecommand \@ifx [1]{%
 \ifx #1\expandafter \@firstoftwo
 \else \expandafter \@secondoftwo
 \fi
}%
\providecommand \natexlab [1]{#1}%
\providecommand \enquote  [1]{``#1''}%
\providecommand \bibnamefont  [1]{#1}%
\providecommand \bibfnamefont [1]{#1}%
\providecommand \citenamefont [1]{#1}%
\providecommand \href@noop [0]{\@secondoftwo}%
\providecommand \href [0]{\begingroup \@sanitize@url \@href}%
\providecommand \@href[1]{\@@startlink{#1}\@@href}%
\providecommand \@@href[1]{\endgroup#1\@@endlink}%
\providecommand \@sanitize@url [0]{\catcode `\\12\catcode `\$12\catcode
  `\&12\catcode `\#12\catcode `\^12\catcode `\_12\catcode `\%12\relax}%
\providecommand \@@startlink[1]{}%
\providecommand \@@endlink[0]{}%
\providecommand \url  [0]{\begingroup\@sanitize@url \@url }%
\providecommand \@url [1]{\endgroup\@href {#1}{\urlprefix }}%
\providecommand \urlprefix  [0]{URL }%
\providecommand \Eprint [0]{\href }%
\providecommand \doibase [0]{https://doi.org/}%
\providecommand \selectlanguage [0]{\@gobble}%
\providecommand \bibinfo  [0]{\@secondoftwo}%
\providecommand \bibfield  [0]{\@secondoftwo}%
\providecommand \translation [1]{[#1]}%
\providecommand \BibitemOpen [0]{}%
\providecommand \bibitemStop [0]{}%
\providecommand \bibitemNoStop [0]{.\EOS\space}%
\providecommand \EOS [0]{\spacefactor3000\relax}%
\providecommand \BibitemShut  [1]{\csname bibitem#1\endcsname}%
\let\auto@bib@innerbib\@empty
\bibitem [{\citenamefont {Damascelli}\ \emph {et~al.}(2003)\citenamefont
  {Damascelli}, \citenamefont {Hussain},\ and\ \citenamefont
  {Shen}}]{Damascelli2003}%
  \BibitemOpen
  \bibfield  {author} {\bibinfo {author} {\bibfnamefont {A.}~\bibnamefont
  {Damascelli}}, \bibinfo {author} {\bibfnamefont {Z.}~\bibnamefont
  {Hussain}},\ and\ \bibinfo {author} {\bibfnamefont {Z.-X.}\ \bibnamefont
  {Shen}},\ }\href {https://doi.org/10.1103/RevModPhys.75.473} {\bibfield
  {journal} {\bibinfo  {journal} {Rev. Mod. Phys.}\ }\textbf {\bibinfo {volume}
  {75}},\ \bibinfo {pages} {473} (\bibinfo {year} {2003})}\BibitemShut
  {NoStop}%
\bibitem [{\citenamefont {Plakida}(2016)}]{Plakida2016}%
  \BibitemOpen
  \bibfield  {author} {\bibinfo {author} {\bibfnamefont {N.~M.}\ \bibnamefont
  {Plakida}},\ }\href
  {https://www.sciencedirect.com/science/article/pii/S092145341630154X}
  {\bibfield  {journal} {\bibinfo  {journal} {Physica C: Superconductivity and
  its Applications}\ }\textbf {\bibinfo {volume} {531}},\ \bibinfo {pages} {39}
  (\bibinfo {year} {2016})}\BibitemShut {NoStop}%
\bibitem [{\citenamefont {Tohyama}\ \emph {et~al.}(1995)\citenamefont
  {Tohyama}, \citenamefont {Horsch},\ and\ \citenamefont
  {Maekawa}}]{Tohyama1995}%
  \BibitemOpen
  \bibfield  {author} {\bibinfo {author} {\bibfnamefont {T.}~\bibnamefont
  {Tohyama}}, \bibinfo {author} {\bibfnamefont {P.}~\bibnamefont {Horsch}},\
  and\ \bibinfo {author} {\bibfnamefont {S.}~\bibnamefont {Maekawa}},\ }\href
  {https://link.aps.org/doi/10.1103/PhysRevLett.74.980} {\bibfield  {journal}
  {\bibinfo  {journal} {PRL}\ }\textbf {\bibinfo {volume} {74}},\ \bibinfo
  {pages} {980} (\bibinfo {year} {1995})}\BibitemShut {NoStop}%
\bibitem [{\citenamefont {Khaliullin}\ and\ \citenamefont
  {Horsch}(1996)}]{Khaliullin1996}%
  \BibitemOpen
  \bibfield  {author} {\bibinfo {author} {\bibfnamefont {G.}~\bibnamefont
  {Khaliullin}}\ and\ \bibinfo {author} {\bibfnamefont {P.}~\bibnamefont
  {Horsch}},\ }\href {https://link.aps.org/doi/10.1103/PhysRevB.54.R9600}
  {\bibfield  {journal} {\bibinfo  {journal} {PRB}\ }\textbf {\bibinfo {volume}
  {54}},\ \bibinfo {pages} {R9600} (\bibinfo {year} {1996})}\BibitemShut
  {NoStop}%
\bibitem [{\citenamefont {Eder}\ \emph {et~al.}(1995)\citenamefont {Eder},
  \citenamefont {Ohta},\ and\ \citenamefont {Maekawa}}]{Eder1995}%
  \BibitemOpen
  \bibfield  {author} {\bibinfo {author} {\bibfnamefont {R.}~\bibnamefont
  {Eder}}, \bibinfo {author} {\bibfnamefont {Y.}~\bibnamefont {Ohta}},\ and\
  \bibinfo {author} {\bibfnamefont {S.}~\bibnamefont {Maekawa}},\ }\href
  {https://link.aps.org/doi/10.1103/PhysRevLett.74.5124} {\bibfield  {journal}
  {\bibinfo  {journal} {PRL}\ }\textbf {\bibinfo {volume} {74}},\ \bibinfo
  {pages} {5124} (\bibinfo {year} {1995})}\BibitemShut {NoStop}%
\bibitem [{\citenamefont {Romero-Bermúdez}\ \emph {et~al.}(2019)\citenamefont
  {Romero-Bermúdez}, \citenamefont {Krikun}, \citenamefont {Schalm},\ and\
  \citenamefont {Zaanen}}]{Romero-Bermudez2019}%
  \BibitemOpen
  \bibfield  {author} {\bibinfo {author} {\bibfnamefont {A.}~\bibnamefont
  {Romero-Bermúdez}}, \bibinfo {author} {\bibfnamefont {A.}~\bibnamefont
  {Krikun}}, \bibinfo {author} {\bibfnamefont {K.}~\bibnamefont {Schalm}},\
  and\ \bibinfo {author} {\bibfnamefont {J.}~\bibnamefont {Zaanen}},\ }\href
  {https://link.aps.org/doi/10.1103/PhysRevB.99.235149} {\bibfield  {journal}
  {\bibinfo  {journal} {PRB}\ }\textbf {\bibinfo {volume} {99}},\ \bibinfo
  {pages} {235149} (\bibinfo {year} {2019})}\BibitemShut {NoStop}%
\bibitem [{\citenamefont {Mahan}(2000)}]{Mahan2000}%
  \BibitemOpen
  \bibfield  {author} {\bibinfo {author} {\bibfnamefont {G.~D.}\ \bibnamefont
  {Mahan}},\ }\href@noop {} {\emph {\bibinfo {title} {Many-Particle Physics}}}\
  (\bibinfo  {publisher} {Kluwer Academic/Plenum Publishers, New York},\
  \bibinfo {year} {2000})\BibitemShut {NoStop}%
\bibitem [{\citenamefont {Vig}\ \emph {et~al.}(2017)\citenamefont {Vig},
  \citenamefont {Kogar}, \citenamefont {Mitrano}, \citenamefont {Husain},
  \citenamefont {Venema}, \citenamefont {Rak}, \citenamefont {Mishra},
  \citenamefont {Johnson}, \citenamefont {Gu}, \citenamefont {Fradkin},
  \citenamefont {Norman},\ and\ \citenamefont {Abbamonte}}]{Vig2017}%
  \BibitemOpen
  \bibfield  {author} {\bibinfo {author} {\bibfnamefont {S.}~\bibnamefont
  {Vig}}, \bibinfo {author} {\bibfnamefont {A.}~\bibnamefont {Kogar}}, \bibinfo
  {author} {\bibfnamefont {M.}~\bibnamefont {Mitrano}}, \bibinfo {author}
  {\bibfnamefont {A.}~\bibnamefont {Husain}}, \bibinfo {author} {\bibfnamefont
  {L.}~\bibnamefont {Venema}}, \bibinfo {author} {\bibfnamefont
  {M.}~\bibnamefont {Rak}}, \bibinfo {author} {\bibfnamefont {V.}~\bibnamefont
  {Mishra}}, \bibinfo {author} {\bibfnamefont {P.}~\bibnamefont {Johnson}},
  \bibinfo {author} {\bibfnamefont {G.}~\bibnamefont {Gu}}, \bibinfo {author}
  {\bibfnamefont {E.}~\bibnamefont {Fradkin}}, \bibinfo {author} {\bibfnamefont
  {M.}~\bibnamefont {Norman}},\ and\ \bibinfo {author} {\bibfnamefont
  {P.}~\bibnamefont {Abbamonte}},\ }\href
  {https://scipost.org/SciPostPhys.3.4.026} {\bibfield  {journal} {\bibinfo
  {journal} {SciPost Phys.}\ }\textbf {\bibinfo {volume} {3}},\ \bibinfo
  {pages} {026} (\bibinfo {year} {2017})}\BibitemShut {NoStop}%
\bibitem [{\citenamefont {Mitrano}\ \emph {et~al.}(2018)\citenamefont
  {Mitrano}, \citenamefont {Husain}, \citenamefont {Vig}, \citenamefont
  {Kogar}, \citenamefont {Rak}, \citenamefont {Rubeck}, \citenamefont
  {Schmalian}, \citenamefont {Uchoa}, \citenamefont {Schneeloch}, \citenamefont
  {Zhong}, \citenamefont {Gu},\ and\ \citenamefont {Abbamonte}}]{Mitrano2018}%
  \BibitemOpen
  \bibfield  {author} {\bibinfo {author} {\bibfnamefont {M.}~\bibnamefont
  {Mitrano}}, \bibinfo {author} {\bibfnamefont {A.~A.}\ \bibnamefont {Husain}},
  \bibinfo {author} {\bibfnamefont {S.}~\bibnamefont {Vig}}, \bibinfo {author}
  {\bibfnamefont {A.}~\bibnamefont {Kogar}}, \bibinfo {author} {\bibfnamefont
  {M.~S.}\ \bibnamefont {Rak}}, \bibinfo {author} {\bibfnamefont {S.~I.}\
  \bibnamefont {Rubeck}}, \bibinfo {author} {\bibfnamefont {J.}~\bibnamefont
  {Schmalian}}, \bibinfo {author} {\bibfnamefont {B.}~\bibnamefont {Uchoa}},
  \bibinfo {author} {\bibfnamefont {J.}~\bibnamefont {Schneeloch}}, \bibinfo
  {author} {\bibfnamefont {R.}~\bibnamefont {Zhong}}, \bibinfo {author}
  {\bibfnamefont {G.~D.}\ \bibnamefont {Gu}},\ and\ \bibinfo {author}
  {\bibfnamefont {P.}~\bibnamefont {Abbamonte}},\ }\href
  {http://www.pnas.org/content/115/21/5392.abstract} {\bibfield  {journal}
  {\bibinfo  {journal} {Proc Natl Acad Sci USA}\ }\textbf {\bibinfo {volume}
  {115}},\ \bibinfo {pages} {5392} (\bibinfo {year} {2018})}\BibitemShut
  {NoStop}%
\bibitem [{\citenamefont {Husain}\ \emph {et~al.}(2020)\citenamefont {Husain},
  \citenamefont {Mitrano}, \citenamefont {Rak}, \citenamefont {Rubeck},
  \citenamefont {Yang}, \citenamefont {Sow}, \citenamefont {Maeno},
  \citenamefont {Batson},\ and\ \citenamefont {Abbamonte}}]{Husain2020}%
  \BibitemOpen
  \bibfield  {author} {\bibinfo {author} {\bibfnamefont {A.~A.}\ \bibnamefont
  {Husain}}, \bibinfo {author} {\bibfnamefont {M.}~\bibnamefont {Mitrano}},
  \bibinfo {author} {\bibfnamefont {M.~S.}\ \bibnamefont {Rak}}, \bibinfo
  {author} {\bibfnamefont {S.~I.}\ \bibnamefont {Rubeck}}, \bibinfo {author}
  {\bibfnamefont {H.}~\bibnamefont {Yang}}, \bibinfo {author} {\bibfnamefont
  {C.}~\bibnamefont {Sow}}, \bibinfo {author} {\bibfnamefont {Y.}~\bibnamefont
  {Maeno}}, \bibinfo {author} {\bibfnamefont {P.~E.}\ \bibnamefont {Batson}},\
  and\ \bibinfo {author} {\bibfnamefont {P.}~\bibnamefont {Abbamonte}},\
  }\href@noop {} {\bibfield  {journal} {\bibinfo  {journal} {arXive
  2007.06670}\ } (\bibinfo {year} {2020})}\BibitemShut {NoStop}%
\bibitem [{\citenamefont {Raether}(1965)}]{Raether1965}%
  \BibitemOpen
  \bibfield  {author} {\bibinfo {author} {\bibfnamefont {H.}~\bibnamefont
  {Raether}},\ }\href@noop {} {\bibfield  {journal} {\bibinfo  {journal}
  {Springer Tracts in Modern Physics}\ }\textbf {\bibinfo {volume} {38}},\
  \bibinfo {pages} {85} (\bibinfo {year} {1965})}\BibitemShut {NoStop}%
\bibitem [{\citenamefont {Daniels}\ \emph {et~al.}(1970)\citenamefont
  {Daniels}, \citenamefont {von Festenberg}, \citenamefont {Raether},\ and\
  \citenamefont {Zeppenfeld}}]{Daniels1970}%
  \BibitemOpen
  \bibfield  {author} {\bibinfo {author} {\bibfnamefont {J.}~\bibnamefont
  {Daniels}}, \bibinfo {author} {\bibfnamefont {C.}~\bibnamefont {von
  Festenberg}}, \bibinfo {author} {\bibfnamefont {H.}~\bibnamefont {Raether}},\
  and\ \bibinfo {author} {\bibfnamefont {K.}~\bibnamefont {Zeppenfeld}},\
  }\href@noop {} {\bibfield  {journal} {\bibinfo  {journal} {Springer Tracts in
  Modern Physics}\ }\textbf {\bibinfo {volume} {54}},\ \bibinfo {pages} {77}
  (\bibinfo {year} {1970})}\BibitemShut {NoStop}%
\bibitem [{\citenamefont {Schnatterly}(1979)}]{Schnatterly1979}%
  \BibitemOpen
  \bibfield  {author} {\bibinfo {author} {\bibfnamefont {J.}~\bibnamefont
  {Schnatterly}},\ }\href@noop {} {\bibfield  {journal} {\bibinfo  {journal}
  {Solid State Phys.}\ }\textbf {\bibinfo {volume} {34}},\ \bibinfo {pages}
  {275} (\bibinfo {year} {1979})}\BibitemShut {NoStop}%
\bibitem [{\citenamefont {Schattschneider}(1986)}]{Schattschneider1986}%
  \BibitemOpen
  \bibfield  {author} {\bibinfo {author} {\bibfnamefont {P.}~\bibnamefont
  {Schattschneider}},\ }\href@noop {} {\emph {\bibinfo {title} {Fundamentals of
  Inelastic Electron Scattering}}}\ (\bibinfo  {publisher} {Springer Verlag,
  Wien},\ \bibinfo {year} {1986})\BibitemShut {NoStop}%
\bibitem [{\citenamefont {Fink}(1989)}]{Fink1989}%
  \BibitemOpen
  \bibfield  {author} {\bibinfo {author} {\bibfnamefont {J.}~\bibnamefont
  {Fink}},\ }\href@noop {} {\bibfield  {journal} {\bibinfo  {journal} {Adv.
  Electr. Electr. Phys.}\ }\textbf {\bibinfo {volume} {75}},\ \bibinfo {pages}
  {121} (\bibinfo {year} {1989})}\BibitemShut {NoStop}%
\bibitem [{\citenamefont {Roth}\ \emph {et~al.}(2014)\citenamefont {Roth},
  \citenamefont {König}, \citenamefont {Fink}, \citenamefont {Büchner},\ and\
  \citenamefont {Knupfer}}]{Roth2014}%
  \BibitemOpen
  \bibfield  {author} {\bibinfo {author} {\bibfnamefont {F.}~\bibnamefont
  {Roth}}, \bibinfo {author} {\bibfnamefont {A.}~\bibnamefont {König}},
  \bibinfo {author} {\bibfnamefont {J.}~\bibnamefont {Fink}}, \bibinfo {author}
  {\bibfnamefont {B.}~\bibnamefont {Büchner}},\ and\ \bibinfo {author}
  {\bibfnamefont {M.}~\bibnamefont {Knupfer}},\ }\href
  {https://www.sciencedirect.com/science/article/pii/S0368204814001169}
  {\bibfield  {journal} {\bibinfo  {journal} {Journal of Electron Spectroscopy
  and Related Phenomena}\ }\textbf {\bibinfo {volume} {195}},\ \bibinfo {pages}
  {85} (\bibinfo {year} {2014})}\BibitemShut {NoStop}%
\bibitem [{\citenamefont {Nücker}\ \emph {et~al.}(1989)\citenamefont
  {Nücker}, \citenamefont {Romberg}, \citenamefont {Nakai}, \citenamefont
  {Scheerer}, \citenamefont {Fink}, \citenamefont {Yan},\ and\ \citenamefont
  {Zhao}}]{Nuecker1989}%
  \BibitemOpen
  \bibfield  {author} {\bibinfo {author} {\bibfnamefont {N.}~\bibnamefont
  {Nücker}}, \bibinfo {author} {\bibfnamefont {H.}~\bibnamefont {Romberg}},
  \bibinfo {author} {\bibfnamefont {S.}~\bibnamefont {Nakai}}, \bibinfo
  {author} {\bibfnamefont {B.}~\bibnamefont {Scheerer}}, \bibinfo {author}
  {\bibfnamefont {J.}~\bibnamefont {Fink}}, \bibinfo {author} {\bibfnamefont
  {Y.~F.}\ \bibnamefont {Yan}},\ and\ \bibinfo {author} {\bibfnamefont {Z.~X.}\
  \bibnamefont {Zhao}},\ }\href
  {https://link.aps.org/doi/10.1103/PhysRevB.39.12379} {\bibfield  {journal}
  {\bibinfo  {journal} {PRB}\ }\textbf {\bibinfo {volume} {39}},\ \bibinfo
  {pages} {12379} (\bibinfo {year} {1989})}\BibitemShut {NoStop}%
\bibitem [{\citenamefont {N{\"u}cker}\ \emph {et~al.}(1991)\citenamefont
  {N{\"u}cker}, \citenamefont {Eckern}, \citenamefont {Fink},\ and\
  \citenamefont {M{\"u}ller}}]{Nuecker1991}%
  \BibitemOpen
  \bibfield  {author} {\bibinfo {author} {\bibfnamefont {N.}~\bibnamefont
  {N{\"u}cker}}, \bibinfo {author} {\bibfnamefont {U.}~\bibnamefont {Eckern}},
  \bibinfo {author} {\bibfnamefont {J.}~\bibnamefont {Fink}},\ and\ \bibinfo
  {author} {\bibfnamefont {P.}~\bibnamefont {M{\"u}ller}},\ }\href
  {https://doi.org/10.1103/PhysRevB.44.7155} {\bibfield  {journal} {\bibinfo
  {journal} {Phys. Rev. B}\ }\textbf {\bibinfo {volume} {44}},\ \bibinfo
  {pages} {7155} (\bibinfo {year} {1991})}\BibitemShut {NoStop}%
\bibitem [{\citenamefont {Vielsack}\ and\ \citenamefont {von
  Baltz}(1990)}]{Vielsack1990}%
  \BibitemOpen
  \bibfield  {author} {\bibinfo {author} {\bibfnamefont {G.}~\bibnamefont
  {Vielsack}}\ and\ \bibinfo {author} {\bibfnamefont {R.}~\bibnamefont {von
  Baltz}},\ }\href {https://doi.org/10.1002/pssb.2221580124} {\bibfield
  {journal} {\bibinfo  {journal} {phys. stat. sol. (b)}\ }\textbf {\bibinfo
  {volume} {158}},\ \bibinfo {pages} {249} (\bibinfo {year}
  {1990})}\BibitemShut {NoStop}%
\bibitem [{\citenamefont {Grigoryan}\ \emph {et~al.}(1999)\citenamefont
  {Grigoryan}, \citenamefont {Paasch},\ and\ \citenamefont
  {Drechsler}}]{Grigoryan1999}%
  \BibitemOpen
  \bibfield  {author} {\bibinfo {author} {\bibfnamefont {V.~G.}\ \bibnamefont
  {Grigoryan}}, \bibinfo {author} {\bibfnamefont {G.}~\bibnamefont {Paasch}},\
  and\ \bibinfo {author} {\bibfnamefont {S.-L.}\ \bibnamefont {Drechsler}},\
  }\href {https://link.aps.org/doi/10.1103/PhysRevB.60.1340} {\bibfield
  {journal} {\bibinfo  {journal} {PRB}\ }\textbf {\bibinfo {volume} {60}},\
  \bibinfo {pages} {1340} (\bibinfo {year} {1999})}\BibitemShut {NoStop}%
\bibitem [{\citenamefont {Paasch}\ and\ \citenamefont
  {Grigoryan}(1999)}]{Paasch1999}%
  \BibitemOpen
  \bibfield  {author} {\bibinfo {author} {\bibfnamefont {G.}~\bibnamefont
  {Paasch}}\ and\ \bibinfo {author} {\bibfnamefont {V.}~\bibnamefont
  {Grigoryan}},\ }\href@noop {} {\bibfield  {journal} {\bibinfo  {journal}
  {Ukr. J. Phys.}\ }\textbf {\bibinfo {volume} {44}},\ \bibinfo {pages} {1480}
  (\bibinfo {year} {1999})}\BibitemShut {NoStop}%
\bibitem [{\citenamefont {Roth}\ \emph {et~al.}(2020)\citenamefont {Roth},
  \citenamefont {Revcolevschi}, \citenamefont {Büchner}, \citenamefont
  {Knupfer},\ and\ \citenamefont {Fink}}]{Roth2020}%
  \BibitemOpen
  \bibfield  {author} {\bibinfo {author} {\bibfnamefont {F.}~\bibnamefont
  {Roth}}, \bibinfo {author} {\bibfnamefont {A.}~\bibnamefont {Revcolevschi}},
  \bibinfo {author} {\bibfnamefont {B.}~\bibnamefont {Büchner}}, \bibinfo
  {author} {\bibfnamefont {M.}~\bibnamefont {Knupfer}},\ and\ \bibinfo {author}
  {\bibfnamefont {J.}~\bibnamefont {Fink}},\ }\href
  {https://link.aps.org/doi/10.1103/PhysRevB.101.195132} {\bibfield  {journal}
  {\bibinfo  {journal} {PRB}\ }\textbf {\bibinfo {volume} {101}},\ \bibinfo
  {pages} {195132} (\bibinfo {year} {2020})}\BibitemShut {NoStop}%
\bibitem [{\citenamefont {Maeno}\ \emph {et~al.}(1994)\citenamefont {Maeno},
  \citenamefont {Hashimoto}, \citenamefont {Yoshida}, \citenamefont
  {Nishizaki}, \citenamefont {Fujita}, \citenamefont {Bednorz},\ and\
  \citenamefont {Lichtenberg}}]{Maeno1994}%
  \BibitemOpen
  \bibfield  {author} {\bibinfo {author} {\bibfnamefont {Y.}~\bibnamefont
  {Maeno}}, \bibinfo {author} {\bibfnamefont {H.}~\bibnamefont {Hashimoto}},
  \bibinfo {author} {\bibfnamefont {K.}~\bibnamefont {Yoshida}}, \bibinfo
  {author} {\bibfnamefont {S.}~\bibnamefont {Nishizaki}}, \bibinfo {author}
  {\bibfnamefont {T.}~\bibnamefont {Fujita}}, \bibinfo {author} {\bibfnamefont
  {J.~G.}\ \bibnamefont {Bednorz}},\ and\ \bibinfo {author} {\bibfnamefont
  {F.}~\bibnamefont {Lichtenberg}},\ }\href {https://doi.org/10.1038/372532a0}
  {\bibfield  {journal} {\bibinfo  {journal} {Nature}\ }\textbf {\bibinfo
  {volume} {372}},\ \bibinfo {pages} {532} (\bibinfo {year}
  {1994})}\BibitemShut {NoStop}%
\bibitem [{\citenamefont {Mackenzie}\ and\ \citenamefont
  {Maeno}(2003)}]{Mackenzie2003}%
  \BibitemOpen
  \bibfield  {author} {\bibinfo {author} {\bibfnamefont {A.~P.}\ \bibnamefont
  {Mackenzie}}\ and\ \bibinfo {author} {\bibfnamefont {Y.}~\bibnamefont
  {Maeno}},\ }\href {https://link.aps.org/doi/10.1103/RevModPhys.75.657}
  {\bibfield  {journal} {\bibinfo  {journal} {RMP}\ }\textbf {\bibinfo {volume}
  {75}},\ \bibinfo {pages} {657} (\bibinfo {year} {2003})}\BibitemShut
  {NoStop}%
\bibitem [{\citenamefont {Sharma}\ \emph {et~al.}(2020)\citenamefont {Sharma},
  \citenamefont {Edkins}, \citenamefont {Wang}, \citenamefont {Kostin},
  \citenamefont {Sow}, \citenamefont {Maeno}, \citenamefont {Mackenzie},
  \citenamefont {Davis},\ and\ \citenamefont {Madhavan}}]{Sharma2020}%
  \BibitemOpen
  \bibfield  {author} {\bibinfo {author} {\bibfnamefont {R.}~\bibnamefont
  {Sharma}}, \bibinfo {author} {\bibfnamefont {S.~D.}\ \bibnamefont {Edkins}},
  \bibinfo {author} {\bibfnamefont {Z.}~\bibnamefont {Wang}}, \bibinfo {author}
  {\bibfnamefont {A.}~\bibnamefont {Kostin}}, \bibinfo {author} {\bibfnamefont
  {C.}~\bibnamefont {Sow}}, \bibinfo {author} {\bibfnamefont {Y.}~\bibnamefont
  {Maeno}}, \bibinfo {author} {\bibfnamefont {A.~P.}\ \bibnamefont
  {Mackenzie}}, \bibinfo {author} {\bibfnamefont {J.~C.~S.}\ \bibnamefont
  {Davis}},\ and\ \bibinfo {author} {\bibfnamefont {V.}~\bibnamefont
  {Madhavan}},\ }\href {https://doi.org/10.1073/pnas.1916463117} {\bibfield
  {journal} {\bibinfo  {journal} {Proceedings of the National Academy of
  Sciences}\ }\textbf {\bibinfo {volume} {117}},\ \bibinfo {pages} {5222}
  (\bibinfo {year} {2020})}\BibitemShut {NoStop}%
\bibitem [{\citenamefont {Katsufuji}\ \emph {et~al.}(1996)\citenamefont
  {Katsufuji}, \citenamefont {Kasai},\ and\ \citenamefont
  {Tokura}}]{Katsufuji1996}%
  \BibitemOpen
  \bibfield  {author} {\bibinfo {author} {\bibfnamefont {T.}~\bibnamefont
  {Katsufuji}}, \bibinfo {author} {\bibfnamefont {M.}~\bibnamefont {Kasai}},\
  and\ \bibinfo {author} {\bibfnamefont {Y.}~\bibnamefont {Tokura}},\ }\href
  {https://link.aps.org/doi/10.1103/PhysRevLett.76.126} {\bibfield  {journal}
  {\bibinfo  {journal} {PRL}\ }\textbf {\bibinfo {volume} {76}},\ \bibinfo
  {pages} {126} (\bibinfo {year} {1996})}\BibitemShut {NoStop}%
\bibitem [{\citenamefont {Hildebrand}\ \emph {et~al.}(2001)\citenamefont
  {Hildebrand}, \citenamefont {Reedyk}, \citenamefont {Katsufuji},\ and\
  \citenamefont {Tokura}}]{Hildebrand2001}%
  \BibitemOpen
  \bibfield  {author} {\bibinfo {author} {\bibfnamefont {M.~G.}\ \bibnamefont
  {Hildebrand}}, \bibinfo {author} {\bibfnamefont {M.}~\bibnamefont {Reedyk}},
  \bibinfo {author} {\bibfnamefont {T.}~\bibnamefont {Katsufuji}},\ and\
  \bibinfo {author} {\bibfnamefont {Y.}~\bibnamefont {Tokura}},\ }\href
  {https://link.aps.org/doi/10.1103/PhysRevLett.87.227002} {\bibfield
  {journal} {\bibinfo  {journal} {PRL}\ }\textbf {\bibinfo {volume} {87}},\
  \bibinfo {pages} {227002} (\bibinfo {year} {2001})}\BibitemShut {NoStop}%
\bibitem [{\citenamefont {Pucher}\ \emph {et~al.}(2003)\citenamefont {Pucher},
  \citenamefont {Loidl}, \citenamefont {Kikugawa},\ and\ \citenamefont
  {Maeno}}]{Pucher2003}%
  \BibitemOpen
  \bibfield  {author} {\bibinfo {author} {\bibfnamefont {K.}~\bibnamefont
  {Pucher}}, \bibinfo {author} {\bibfnamefont {A.}~\bibnamefont {Loidl}},
  \bibinfo {author} {\bibfnamefont {N.}~\bibnamefont {Kikugawa}},\ and\
  \bibinfo {author} {\bibfnamefont {Y.}~\bibnamefont {Maeno}},\ }\href
  {https://link.aps.org/doi/10.1103/PhysRevB.68.214502} {\bibfield  {journal}
  {\bibinfo  {journal} {PRB}\ }\textbf {\bibinfo {volume} {68}},\ \bibinfo
  {pages} {214502} (\bibinfo {year} {2003})}\BibitemShut {NoStop}%
\bibitem [{\citenamefont {Stricker}\ \emph {et~al.}(2014)\citenamefont
  {Stricker}, \citenamefont {Mravlje}, \citenamefont {Berthod}, \citenamefont
  {Fittipaldi}, \citenamefont {Vecchione}, \citenamefont {Georges},\ and\
  \citenamefont {van~der Marel}}]{Stricker2014}%
  \BibitemOpen
  \bibfield  {author} {\bibinfo {author} {\bibfnamefont {D.}~\bibnamefont
  {Stricker}}, \bibinfo {author} {\bibfnamefont {J.}~\bibnamefont {Mravlje}},
  \bibinfo {author} {\bibfnamefont {C.}~\bibnamefont {Berthod}}, \bibinfo
  {author} {\bibfnamefont {R.}~\bibnamefont {Fittipaldi}}, \bibinfo {author}
  {\bibfnamefont {A.}~\bibnamefont {Vecchione}}, \bibinfo {author}
  {\bibfnamefont {A.}~\bibnamefont {Georges}},\ and\ \bibinfo {author}
  {\bibfnamefont {D.}~\bibnamefont {van~der Marel}},\ }\href
  {https://link.aps.org/doi/10.1103/PhysRevLett.113.087404} {\bibfield
  {journal} {\bibinfo  {journal} {PRL}\ }\textbf {\bibinfo {volume} {113}},\
  \bibinfo {pages} {087404} (\bibinfo {year} {2014})}\BibitemShut {NoStop}%
\bibitem [{\citenamefont {Stricker}(2015)}]{Stricker2015}%
  \BibitemOpen
  \bibfield  {author} {\bibinfo {author} {\bibfnamefont {D.}~\bibnamefont
  {Stricker}},\ }\href@noop {} {Ph.D. thesis},\ \bibinfo  {school} {University
  Geneve} (\bibinfo {year} {2015})\BibitemShut {NoStop}%
\bibitem [{\citenamefont {Wang}\ \emph {et~al.}(2017)\citenamefont {Wang},
  \citenamefont {Walkup}, \citenamefont {Derry}, \citenamefont {Scaffidi},
  \citenamefont {Rak}, \citenamefont {Vig}, \citenamefont {Kogar},
  \citenamefont {Zeljkovic}, \citenamefont {Husain}, \citenamefont {Santos},
  \citenamefont {Wang}, \citenamefont {Damascelli}, \citenamefont {Maeno},
  \citenamefont {Abbamonte}, \citenamefont {Fradkin},\ and\ \citenamefont
  {Madhavan}}]{Wang2017}%
  \BibitemOpen
  \bibfield  {author} {\bibinfo {author} {\bibfnamefont {Z.}~\bibnamefont
  {Wang}}, \bibinfo {author} {\bibfnamefont {D.}~\bibnamefont {Walkup}},
  \bibinfo {author} {\bibfnamefont {P.}~\bibnamefont {Derry}}, \bibinfo
  {author} {\bibfnamefont {T.}~\bibnamefont {Scaffidi}}, \bibinfo {author}
  {\bibfnamefont {M.}~\bibnamefont {Rak}}, \bibinfo {author} {\bibfnamefont
  {S.}~\bibnamefont {Vig}}, \bibinfo {author} {\bibfnamefont {A.}~\bibnamefont
  {Kogar}}, \bibinfo {author} {\bibfnamefont {I.}~\bibnamefont {Zeljkovic}},
  \bibinfo {author} {\bibfnamefont {A.}~\bibnamefont {Husain}}, \bibinfo
  {author} {\bibfnamefont {L.~H.}\ \bibnamefont {Santos}}, \bibinfo {author}
  {\bibfnamefont {Y.}~\bibnamefont {Wang}}, \bibinfo {author} {\bibfnamefont
  {A.}~\bibnamefont {Damascelli}}, \bibinfo {author} {\bibfnamefont
  {Y.}~\bibnamefont {Maeno}}, \bibinfo {author} {\bibfnamefont
  {P.}~\bibnamefont {Abbamonte}}, \bibinfo {author} {\bibfnamefont
  {E.}~\bibnamefont {Fradkin}},\ and\ \bibinfo {author} {\bibfnamefont
  {V.}~\bibnamefont {Madhavan}},\ }\href {https://doi.org/10.1038/nphys4107}
  {\bibfield  {journal} {\bibinfo  {journal} {Nature Physics}\ }\textbf
  {\bibinfo {volume} {13}},\ \bibinfo {pages} {799} (\bibinfo {year}
  {2017})}\BibitemShut {NoStop}%
\bibitem [{\citenamefont {Wang}\ \emph {et~al.}(2004)\citenamefont {Wang},
  \citenamefont {Yang}, \citenamefont {Sekharan}, \citenamefont {Ding},
  \citenamefont {Engelbrecht}, \citenamefont {Dai}, \citenamefont {Wang},
  \citenamefont {Kaminski}, \citenamefont {Valla}, \citenamefont {Kidd},
  \citenamefont {Fedorov},\ and\ \citenamefont {Johnson}}]{Wang2004}%
  \BibitemOpen
  \bibfield  {author} {\bibinfo {author} {\bibfnamefont {S.-C.}\ \bibnamefont
  {Wang}}, \bibinfo {author} {\bibfnamefont {H.-B.}\ \bibnamefont {Yang}},
  \bibinfo {author} {\bibfnamefont {A.~K.~P.}\ \bibnamefont {Sekharan}},
  \bibinfo {author} {\bibfnamefont {H.}~\bibnamefont {Ding}}, \bibinfo {author}
  {\bibfnamefont {J.~R.}\ \bibnamefont {Engelbrecht}}, \bibinfo {author}
  {\bibfnamefont {X.}~\bibnamefont {Dai}}, \bibinfo {author} {\bibfnamefont
  {Z.}~\bibnamefont {Wang}}, \bibinfo {author} {\bibfnamefont {A.}~\bibnamefont
  {Kaminski}}, \bibinfo {author} {\bibfnamefont {T.}~\bibnamefont {Valla}},
  \bibinfo {author} {\bibfnamefont {T.}~\bibnamefont {Kidd}}, \bibinfo {author}
  {\bibfnamefont {A.~V.}\ \bibnamefont {Fedorov}},\ and\ \bibinfo {author}
  {\bibfnamefont {P.~D.}\ \bibnamefont {Johnson}},\ }\href
  {https://link.aps.org/doi/10.1103/PhysRevLett.92.137002} {\bibfield
  {journal} {\bibinfo  {journal} {PRL}\ }\textbf {\bibinfo {volume} {92}},\
  \bibinfo {pages} {137002} (\bibinfo {year} {2004})}\BibitemShut {NoStop}%
\bibitem [{\citenamefont {Ingle}\ \emph {et~al.}(2005)\citenamefont {Ingle},
  \citenamefont {Shen}, \citenamefont {Baumberger}, \citenamefont {Meevasana},
  \citenamefont {Lu}, \citenamefont {Shen}, \citenamefont {Damascelli},
  \citenamefont {Nakatsuji}, \citenamefont {Mao}, \citenamefont {Maeno},
  \citenamefont {Kimura},\ and\ \citenamefont {Tokura}}]{Ingle2005}%
  \BibitemOpen
  \bibfield  {author} {\bibinfo {author} {\bibfnamefont {N.~J.~C.}\
  \bibnamefont {Ingle}}, \bibinfo {author} {\bibfnamefont {K.~M.}\ \bibnamefont
  {Shen}}, \bibinfo {author} {\bibfnamefont {F.}~\bibnamefont {Baumberger}},
  \bibinfo {author} {\bibfnamefont {W.}~\bibnamefont {Meevasana}}, \bibinfo
  {author} {\bibfnamefont {D.~H.}\ \bibnamefont {Lu}}, \bibinfo {author}
  {\bibfnamefont {Z.-X.}\ \bibnamefont {Shen}}, \bibinfo {author}
  {\bibfnamefont {A.}~\bibnamefont {Damascelli}}, \bibinfo {author}
  {\bibfnamefont {S.}~\bibnamefont {Nakatsuji}}, \bibinfo {author}
  {\bibfnamefont {Z.~Q.}\ \bibnamefont {Mao}}, \bibinfo {author} {\bibfnamefont
  {Y.}~\bibnamefont {Maeno}}, \bibinfo {author} {\bibfnamefont
  {T.}~\bibnamefont {Kimura}},\ and\ \bibinfo {author} {\bibfnamefont
  {Y.}~\bibnamefont {Tokura}},\ }\href
  {https://link.aps.org/doi/10.1103/PhysRevB.72.205114} {\bibfield  {journal}
  {\bibinfo  {journal} {PRB}\ }\textbf {\bibinfo {volume} {72}},\ \bibinfo
  {pages} {205114} (\bibinfo {year} {2005})}\BibitemShut {NoStop}%
\bibitem [{\citenamefont {Zabolotnyy}\ \emph {et~al.}(2013)\citenamefont
  {Zabolotnyy}, \citenamefont {Evtushinsky}, \citenamefont {Kordyuk},
  \citenamefont {Kim}, \citenamefont {Carleschi}, \citenamefont {Doyle},
  \citenamefont {Fittipaldi}, \citenamefont {Cuoco}, \citenamefont
  {Vecchione},\ and\ \citenamefont {Borisenko}}]{Zabolotnyy2013}%
  \BibitemOpen
  \bibfield  {author} {\bibinfo {author} {\bibfnamefont {V.~B.}\ \bibnamefont
  {Zabolotnyy}}, \bibinfo {author} {\bibfnamefont {D.~V.}\ \bibnamefont
  {Evtushinsky}}, \bibinfo {author} {\bibfnamefont {A.~A.}\ \bibnamefont
  {Kordyuk}}, \bibinfo {author} {\bibfnamefont {T.~K.}\ \bibnamefont {Kim}},
  \bibinfo {author} {\bibfnamefont {E.}~\bibnamefont {Carleschi}}, \bibinfo
  {author} {\bibfnamefont {B.~P.}\ \bibnamefont {Doyle}}, \bibinfo {author}
  {\bibfnamefont {R.}~\bibnamefont {Fittipaldi}}, \bibinfo {author}
  {\bibfnamefont {M.}~\bibnamefont {Cuoco}}, \bibinfo {author} {\bibfnamefont
  {A.}~\bibnamefont {Vecchione}},\ and\ \bibinfo {author} {\bibfnamefont
  {S.~V.}\ \bibnamefont {Borisenko}},\ }\href
  {https://www.sciencedirect.com/science/article/pii/S0368204813001655}
  {\bibfield  {journal} {\bibinfo  {journal} {Journal of Electron Spectroscopy
  and Related Phenomena}\ }\textbf {\bibinfo {volume} {191}},\ \bibinfo {pages}
  {48} (\bibinfo {year} {2013})}\BibitemShut {NoStop}%
\bibitem [{\citenamefont {Akebi}\ \emph {et~al.}(2019)\citenamefont {Akebi},
  \citenamefont {Kondo}, \citenamefont {Nakayama}, \citenamefont {Kuroda},
  \citenamefont {Kunisada}, \citenamefont {Taniguchi}, \citenamefont {Maeno},\
  and\ \citenamefont {Shin}}]{Akebi2019}%
  \BibitemOpen
  \bibfield  {author} {\bibinfo {author} {\bibfnamefont {S.}~\bibnamefont
  {Akebi}}, \bibinfo {author} {\bibfnamefont {T.}~\bibnamefont {Kondo}},
  \bibinfo {author} {\bibfnamefont {M.}~\bibnamefont {Nakayama}}, \bibinfo
  {author} {\bibfnamefont {K.}~\bibnamefont {Kuroda}}, \bibinfo {author}
  {\bibfnamefont {S.}~\bibnamefont {Kunisada}}, \bibinfo {author}
  {\bibfnamefont {H.}~\bibnamefont {Taniguchi}}, \bibinfo {author}
  {\bibfnamefont {Y.}~\bibnamefont {Maeno}},\ and\ \bibinfo {author}
  {\bibfnamefont {S.}~\bibnamefont {Shin}},\ }\href
  {https://link.aps.org/doi/10.1103/PhysRevB.99.081108} {\bibfield  {journal}
  {\bibinfo  {journal} {PRB}\ }\textbf {\bibinfo {volume} {99}},\ \bibinfo
  {pages} {081108} (\bibinfo {year} {2019})}\BibitemShut {NoStop}%
\bibitem [{\citenamefont {Tamai}\ \emph {et~al.}(2019)\citenamefont {Tamai},
  \citenamefont {Zingl}, \citenamefont {Rozbicki}, \citenamefont {Cappelli},
  \citenamefont {Riccò}, \citenamefont {de~la Torre}, \citenamefont
  {McKeown~Walker}, \citenamefont {Bruno}, \citenamefont {King}, \citenamefont
  {Meevasana}, \citenamefont {Shi}, \citenamefont {Radović}, \citenamefont
  {Plumb}, \citenamefont {Gibbs}, \citenamefont {Mackenzie}, \citenamefont
  {Berthod}, \citenamefont {Strand}, \citenamefont {Kim}, \citenamefont
  {Georges},\ and\ \citenamefont {Baumberger}}]{Tamai2019}%
  \BibitemOpen
  \bibfield  {author} {\bibinfo {author} {\bibfnamefont {A.}~\bibnamefont
  {Tamai}}, \bibinfo {author} {\bibfnamefont {M.}~\bibnamefont {Zingl}},
  \bibinfo {author} {\bibfnamefont {E.}~\bibnamefont {Rozbicki}}, \bibinfo
  {author} {\bibfnamefont {E.}~\bibnamefont {Cappelli}}, \bibinfo {author}
  {\bibfnamefont {S.}~\bibnamefont {Riccò}}, \bibinfo {author} {\bibfnamefont
  {A.}~\bibnamefont {de~la Torre}}, \bibinfo {author} {\bibfnamefont
  {S.}~\bibnamefont {McKeown~Walker}}, \bibinfo {author} {\bibfnamefont
  {F.~Y.}\ \bibnamefont {Bruno}}, \bibinfo {author} {\bibfnamefont {P.~D.~C.}\
  \bibnamefont {King}}, \bibinfo {author} {\bibfnamefont {W.}~\bibnamefont
  {Meevasana}}, \bibinfo {author} {\bibfnamefont {M.}~\bibnamefont {Shi}},
  \bibinfo {author} {\bibfnamefont {M.}~\bibnamefont {Radović}}, \bibinfo
  {author} {\bibfnamefont {N.~C.}\ \bibnamefont {Plumb}}, \bibinfo {author}
  {\bibfnamefont {A.~S.}\ \bibnamefont {Gibbs}}, \bibinfo {author}
  {\bibfnamefont {A.~P.}\ \bibnamefont {Mackenzie}}, \bibinfo {author}
  {\bibfnamefont {C.}~\bibnamefont {Berthod}}, \bibinfo {author} {\bibfnamefont
  {H.~U.~R.}\ \bibnamefont {Strand}}, \bibinfo {author} {\bibfnamefont
  {M.}~\bibnamefont {Kim}}, \bibinfo {author} {\bibfnamefont {A.}~\bibnamefont
  {Georges}},\ and\ \bibinfo {author} {\bibfnamefont {F.}~\bibnamefont
  {Baumberger}},\ }\href {https://link.aps.org/doi/10.1103/PhysRevX.9.021048}
  {\bibfield  {journal} {\bibinfo  {journal} {PRX}\ }\textbf {\bibinfo {volume}
  {9}},\ \bibinfo {pages} {021048} (\bibinfo {year} {2019})}\BibitemShut
  {NoStop}%
\bibitem [{\citenamefont {Singh}(1995{\natexlab{a}})}]{Singh1995}%
  \BibitemOpen
  \bibfield  {author} {\bibinfo {author} {\bibfnamefont {D.~J.}\ \bibnamefont
  {Singh}},\ }\href {https://link.aps.org/doi/10.1103/PhysRevB.52.1358}
  {\bibfield  {journal} {\bibinfo  {journal} {PRB}\ }\textbf {\bibinfo {volume}
  {52}},\ \bibinfo {pages} {1358} (\bibinfo {year}
  {1995}{\natexlab{a}})}\BibitemShut {NoStop}%
\bibitem [{\citenamefont {Liebsch}\ and\ \citenamefont
  {Lichtenstein}(2000)}]{Liebsch2000}%
  \BibitemOpen
  \bibfield  {author} {\bibinfo {author} {\bibfnamefont {A.}~\bibnamefont
  {Liebsch}}\ and\ \bibinfo {author} {\bibfnamefont {A.}~\bibnamefont
  {Lichtenstein}},\ }\href
  {https://link.aps.org/doi/10.1103/PhysRevLett.84.1591} {\bibfield  {journal}
  {\bibinfo  {journal} {PRL}\ }\textbf {\bibinfo {volume} {84}},\ \bibinfo
  {pages} {1591} (\bibinfo {year} {2000})}\BibitemShut {NoStop}%
\bibitem [{\citenamefont {Singh}(1995{\natexlab{b}})}]{Singh_1995}%
  \BibitemOpen
  \bibfield  {author} {\bibinfo {author} {\bibfnamefont {D.~J.}\ \bibnamefont
  {Singh}},\ }\href {https://doi.org/10.1103/physrevb.52.1358} {\bibfield
  {journal} {\bibinfo  {journal} {Physical Review B}\ }\textbf {\bibinfo
  {volume} {52}},\ \bibinfo {pages} {1358} (\bibinfo {year}
  {1995}{\natexlab{b}})}\BibitemShut {NoStop}%
\bibitem [{\citenamefont {Bobowski}\ \emph {et~al.}(2019)\citenamefont
  {Bobowski}, \citenamefont {Kikugawa}, \citenamefont {Miyoshi}, \citenamefont
  {Suwa}, \citenamefont {shu Xu}, \citenamefont {Yonezawa}, \citenamefont
  {Sokolov}, \citenamefont {Mackenzie},\ and\ \citenamefont
  {Maeno}}]{Bobowski2019}%
  \BibitemOpen
  \bibfield  {author} {\bibinfo {author} {\bibfnamefont {J.~S.}\ \bibnamefont
  {Bobowski}}, \bibinfo {author} {\bibfnamefont {N.}~\bibnamefont {Kikugawa}},
  \bibinfo {author} {\bibfnamefont {T.}~\bibnamefont {Miyoshi}}, \bibinfo
  {author} {\bibfnamefont {H.}~\bibnamefont {Suwa}}, \bibinfo {author}
  {\bibfnamefont {H.}~\bibnamefont {shu Xu}}, \bibinfo {author} {\bibfnamefont
  {S.}~\bibnamefont {Yonezawa}}, \bibinfo {author} {\bibfnamefont {D.~A.}\
  \bibnamefont {Sokolov}}, \bibinfo {author} {\bibfnamefont {A.~P.}\
  \bibnamefont {Mackenzie}},\ and\ \bibinfo {author} {\bibfnamefont
  {Y.}~\bibnamefont {Maeno}},\ }\href {https://doi.org/10.3390/condmat4010006}
  {\bibfield  {journal} {\bibinfo  {journal} {Condensed Matter}\ }\textbf
  {\bibinfo {volume} {4}},\ \bibinfo {pages} {6} (\bibinfo {year}
  {2019})}\BibitemShut {NoStop}%
\bibitem [{\citenamefont {Ehrenreich}\ and\ \citenamefont
  {Cohen}(1959)}]{Ehrenreich1959}%
  \BibitemOpen
  \bibfield  {author} {\bibinfo {author} {\bibfnamefont {H.}~\bibnamefont
  {Ehrenreich}}\ and\ \bibinfo {author} {\bibfnamefont {M.~H.}\ \bibnamefont
  {Cohen}},\ }\href {https://link.aps.org/doi/10.1103/PhysRev.115.786}
  {\bibfield  {journal} {\bibinfo  {journal} {Phys. Rev.}\ }\textbf {\bibinfo
  {volume} {115}},\ \bibinfo {pages} {786} (\bibinfo {year}
  {1959})}\BibitemShut {NoStop}%
\bibitem [{\citenamefont {Giuliani}\ and\ \citenamefont
  {Vignale}(2005)}]{Giuliani2005}%
  \BibitemOpen
  \bibfield  {author} {\bibinfo {author} {\bibfnamefont {G.~F.}\ \bibnamefont
  {Giuliani}}\ and\ \bibinfo {author} {\bibfnamefont {G.}~\bibnamefont
  {Vignale}},\ }\href@noop {} {\emph {\bibinfo {title} {Quantum Theory of the
  Electron Liquid}}}\ (\bibinfo  {publisher} {Cambridge University Press},\
  \bibinfo {year} {2005})\BibitemShut {NoStop}%
\bibitem [{\citenamefont {Neumann}\ and\ \citenamefont {von
  Baltz}(1987)}]{Neumann1987}%
  \BibitemOpen
  \bibfield  {author} {\bibinfo {author} {\bibfnamefont {C.-S.}\ \bibnamefont
  {Neumann}}\ and\ \bibinfo {author} {\bibfnamefont {R.}~\bibnamefont {von
  Baltz}},\ }\href {https://link.aps.org/doi/10.1103/PhysRevB.35.9708}
  {\bibfield  {journal} {\bibinfo  {journal} {PRB}\ }\textbf {\bibinfo {volume}
  {35}},\ \bibinfo {pages} {9708} (\bibinfo {year} {1987})}\BibitemShut
  {NoStop}%
\bibitem [{\citenamefont {Pchelkina}\ \emph {et~al.}(2007)\citenamefont
  {Pchelkina}, \citenamefont {Nekrasov}, \citenamefont {Pruschke},
  \citenamefont {Sekiyama}, \citenamefont {Suga}, \citenamefont {Anisimov},\
  and\ \citenamefont {Vollhardt}}]{Pchelkina2007}%
  \BibitemOpen
  \bibfield  {author} {\bibinfo {author} {\bibfnamefont {Z.~V.}\ \bibnamefont
  {Pchelkina}}, \bibinfo {author} {\bibfnamefont {I.~A.}\ \bibnamefont
  {Nekrasov}}, \bibinfo {author} {\bibfnamefont {T.}~\bibnamefont {Pruschke}},
  \bibinfo {author} {\bibfnamefont {A.}~\bibnamefont {Sekiyama}}, \bibinfo
  {author} {\bibfnamefont {S.}~\bibnamefont {Suga}}, \bibinfo {author}
  {\bibfnamefont {V.~I.}\ \bibnamefont {Anisimov}},\ and\ \bibinfo {author}
  {\bibfnamefont {D.}~\bibnamefont {Vollhardt}},\ }\bibfield  {journal}
  {\bibinfo  {journal} {Physical Review B}\ }\textbf {\bibinfo {volume} {75}},\
  \href {https://doi.org/10.1103/physrevb.75.035122}
  {10.1103/physrevb.75.035122} (\bibinfo {year} {2007})\BibitemShut {NoStop}%
\bibitem [{\citenamefont {DuBois}\ and\ \citenamefont
  {Kivelson}(1969)}]{DuBois1969}%
  \BibitemOpen
  \bibfield  {author} {\bibinfo {author} {\bibfnamefont {D.~F.}\ \bibnamefont
  {DuBois}}\ and\ \bibinfo {author} {\bibfnamefont {M.~G.}\ \bibnamefont
  {Kivelson}},\ }\href {https://doi.org/10.1103/physrev.186.409} {\bibfield
  {journal} {\bibinfo  {journal} {Physical Review}\ }\textbf {\bibinfo {volume}
  {186}},\ \bibinfo {pages} {409} (\bibinfo {year} {1969})}\BibitemShut
  {NoStop}%
\bibitem [{\citenamefont {Paasch}(1970)}]{Paasch1970}%
  \BibitemOpen
  \bibfield  {author} {\bibinfo {author} {\bibfnamefont {G.}~\bibnamefont
  {Paasch}},\ }\href@noop {} {\bibfield  {journal} {\bibinfo  {journal} {phys.
  stat. sol.}\ }\textbf {\bibinfo {volume} {38}},\ \bibinfo {pages} {K123}
  (\bibinfo {year} {1970})}\BibitemShut {NoStop}%
\bibitem [{\citenamefont {Gibbons}\ \emph {et~al.}(1976)\citenamefont
  {Gibbons}, \citenamefont {Schnatterly}, \citenamefont {Ritsko},\ and\
  \citenamefont {Fields}}]{Gibbons1976}%
  \BibitemOpen
  \bibfield  {author} {\bibinfo {author} {\bibfnamefont {P.~C.}\ \bibnamefont
  {Gibbons}}, \bibinfo {author} {\bibfnamefont {S.~E.}\ \bibnamefont
  {Schnatterly}}, \bibinfo {author} {\bibfnamefont {J.~J.}\ \bibnamefont
  {Ritsko}},\ and\ \bibinfo {author} {\bibfnamefont {J.~R.}\ \bibnamefont
  {Fields}},\ }\href {https://link.aps.org/doi/10.1103/PhysRevB.13.2451}
  {\bibfield  {journal} {\bibinfo  {journal} {PRB}\ }\textbf {\bibinfo {volume}
  {13}},\ \bibinfo {pages} {2451} (\bibinfo {year} {1976})}\BibitemShut
  {NoStop}%
\bibitem [{\citenamefont {vom Felde}\ \emph
  {et~al.}(1989{\natexlab{a}})\citenamefont {vom Felde}, \citenamefont
  {Sprösser-Prou},\ and\ \citenamefont {Fink}}]{Felde1989}%
  \BibitemOpen
  \bibfield  {author} {\bibinfo {author} {\bibfnamefont {A.}~\bibnamefont {vom
  Felde}}, \bibinfo {author} {\bibfnamefont {J.}~\bibnamefont
  {Sprösser-Prou}},\ and\ \bibinfo {author} {\bibfnamefont {J.}~\bibnamefont
  {Fink}},\ }\href {https://link.aps.org/doi/10.1103/PhysRevB.40.10181}
  {\bibfield  {journal} {\bibinfo  {journal} {PRB}\ }\textbf {\bibinfo {volume}
  {40}},\ \bibinfo {pages} {10181} (\bibinfo {year}
  {1989}{\natexlab{a}})}\BibitemShut {NoStop}%
\bibitem [{\citenamefont {Grecu}(1973)}]{Grecu1973}%
  \BibitemOpen
  \bibfield  {author} {\bibinfo {author} {\bibfnamefont {D.}~\bibnamefont
  {Grecu}},\ }\href {https://link.aps.org/doi/10.1103/PhysRevB.8.1958}
  {\bibfield  {journal} {\bibinfo  {journal} {PRB}\ }\textbf {\bibinfo {volume}
  {8}},\ \bibinfo {pages} {1958} (\bibinfo {year} {1973})}\BibitemShut
  {NoStop}%
\bibitem [{\citenamefont {Fetter}(1974)}]{Fetter1974}%
  \BibitemOpen
  \bibfield  {author} {\bibinfo {author} {\bibfnamefont {A.~L.}\ \bibnamefont
  {Fetter}},\ }\href
  {https://www.sciencedirect.com/science/article/pii/0003491674903972}
  {\bibfield  {journal} {\bibinfo  {journal} {Annals of Physics}\ }\textbf
  {\bibinfo {volume} {88}},\ \bibinfo {pages} {1} (\bibinfo {year}
  {1974})}\BibitemShut {NoStop}%
\bibitem [{\citenamefont {Giuliani}\ and\ \citenamefont
  {Quinn}(1983)}]{Giuliani1983}%
  \BibitemOpen
  \bibfield  {author} {\bibinfo {author} {\bibfnamefont {G.~F.}\ \bibnamefont
  {Giuliani}}\ and\ \bibinfo {author} {\bibfnamefont {J.~J.}\ \bibnamefont
  {Quinn}},\ }\href {https://link.aps.org/doi/10.1103/PhysRevLett.51.919}
  {\bibfield  {journal} {\bibinfo  {journal} {PRL}\ }\textbf {\bibinfo {volume}
  {51}},\ \bibinfo {pages} {919} (\bibinfo {year} {1983})}\BibitemShut
  {NoStop}%
\bibitem [{\citenamefont {Kresin}\ and\ \citenamefont
  {Morawitz}(1988)}]{Kresin1988}%
  \BibitemOpen
  \bibfield  {author} {\bibinfo {author} {\bibfnamefont {V.~Z.}\ \bibnamefont
  {Kresin}}\ and\ \bibinfo {author} {\bibfnamefont {H.}~\bibnamefont
  {Morawitz}},\ }\href {https://link.aps.org/doi/10.1103/PhysRevB.37.7854}
  {\bibfield  {journal} {\bibinfo  {journal} {PRB}\ }\textbf {\bibinfo {volume}
  {37}},\ \bibinfo {pages} {7854} (\bibinfo {year} {1988})}\BibitemShut
  {NoStop}%
\bibitem [{\citenamefont {Markiewicz}\ and\ \citenamefont
  {Bansil}(2007)}]{Markiewicz2007}%
  \BibitemOpen
  \bibfield  {author} {\bibinfo {author} {\bibfnamefont {R.~S.}\ \bibnamefont
  {Markiewicz}}\ and\ \bibinfo {author} {\bibfnamefont {A.}~\bibnamefont
  {Bansil}},\ }\href {https://link.aps.org/doi/10.1103/PhysRevB.75.020508}
  {\bibfield  {journal} {\bibinfo  {journal} {PRB}\ }\textbf {\bibinfo {volume}
  {75}},\ \bibinfo {pages} {020508} (\bibinfo {year} {2007})}\BibitemShut
  {NoStop}%
\bibitem [{\citenamefont {Greco}\ \emph {et~al.}(2016)\citenamefont {Greco},
  \citenamefont {Yamase},\ and\ \citenamefont {Bejas}}]{Greco2016}%
  \BibitemOpen
  \bibfield  {author} {\bibinfo {author} {\bibfnamefont {A.}~\bibnamefont
  {Greco}}, \bibinfo {author} {\bibfnamefont {H.}~\bibnamefont {Yamase}},\ and\
  \bibinfo {author} {\bibfnamefont {M.}~\bibnamefont {Bejas}},\ }\href
  {https://link.aps.org/doi/10.1103/PhysRevB.94.075139} {\bibfield  {journal}
  {\bibinfo  {journal} {PRB}\ }\textbf {\bibinfo {volume} {94}},\ \bibinfo
  {pages} {075139} (\bibinfo {year} {2016})}\BibitemShut {NoStop}%
\bibitem [{\citenamefont {Ishii}\ \emph {et~al.}(2017)\citenamefont {Ishii},
  \citenamefont {Tohyama}, \citenamefont {Asano}, \citenamefont {Sato},
  \citenamefont {Fujita}, \citenamefont {Wakimoto}, \citenamefont {Tustsui},
  \citenamefont {Sota}, \citenamefont {Miyawaki}, \citenamefont {Niwa},
  \citenamefont {Harada}, \citenamefont {Pelliciari}, \citenamefont {Huang},
  \citenamefont {Schmitt}, \citenamefont {Yamamoto},\ and\ \citenamefont
  {Mizuki}}]{Ishii2017}%
  \BibitemOpen
  \bibfield  {author} {\bibinfo {author} {\bibfnamefont {K.}~\bibnamefont
  {Ishii}}, \bibinfo {author} {\bibfnamefont {T.}~\bibnamefont {Tohyama}},
  \bibinfo {author} {\bibfnamefont {S.}~\bibnamefont {Asano}}, \bibinfo
  {author} {\bibfnamefont {K.}~\bibnamefont {Sato}}, \bibinfo {author}
  {\bibfnamefont {M.}~\bibnamefont {Fujita}}, \bibinfo {author} {\bibfnamefont
  {S.}~\bibnamefont {Wakimoto}}, \bibinfo {author} {\bibfnamefont
  {K.}~\bibnamefont {Tustsui}}, \bibinfo {author} {\bibfnamefont
  {S.}~\bibnamefont {Sota}}, \bibinfo {author} {\bibfnamefont {J.}~\bibnamefont
  {Miyawaki}}, \bibinfo {author} {\bibfnamefont {H.}~\bibnamefont {Niwa}},
  \bibinfo {author} {\bibfnamefont {Y.}~\bibnamefont {Harada}}, \bibinfo
  {author} {\bibfnamefont {J.}~\bibnamefont {Pelliciari}}, \bibinfo {author}
  {\bibfnamefont {Y.}~\bibnamefont {Huang}}, \bibinfo {author} {\bibfnamefont
  {T.}~\bibnamefont {Schmitt}}, \bibinfo {author} {\bibfnamefont
  {Y.}~\bibnamefont {Yamamoto}},\ and\ \bibinfo {author} {\bibfnamefont
  {J.}~\bibnamefont {Mizuki}},\ }\href
  {https://link.aps.org/doi/10.1103/PhysRevB.96.115148} {\bibfield  {journal}
  {\bibinfo  {journal} {PRB}\ }\textbf {\bibinfo {volume} {96}},\ \bibinfo
  {pages} {115148} (\bibinfo {year} {2017})}\BibitemShut {NoStop}%
\bibitem [{\citenamefont {Hepting}\ \emph {et~al.}(2018)\citenamefont
  {Hepting}, \citenamefont {Chaix}, \citenamefont {Huang}, \citenamefont
  {Fumagalli}, \citenamefont {Peng}, \citenamefont {Moritz}, \citenamefont
  {Kummer}, \citenamefont {Brookes}, \citenamefont {Lee}, \citenamefont
  {Hashimoto}, \citenamefont {Sarkar}, \citenamefont {He}, \citenamefont
  {Rotundu}, \citenamefont {Lee}, \citenamefont {Greene}, \citenamefont
  {Braicovich}, \citenamefont {Ghiringhelli}, \citenamefont {Shen},
  \citenamefont {Devereaux},\ and\ \citenamefont {Lee}}]{Hepting2018}%
  \BibitemOpen
  \bibfield  {author} {\bibinfo {author} {\bibfnamefont {M.}~\bibnamefont
  {Hepting}}, \bibinfo {author} {\bibfnamefont {L.}~\bibnamefont {Chaix}},
  \bibinfo {author} {\bibfnamefont {E.~W.}\ \bibnamefont {Huang}}, \bibinfo
  {author} {\bibfnamefont {R.}~\bibnamefont {Fumagalli}}, \bibinfo {author}
  {\bibfnamefont {Y.~Y.}\ \bibnamefont {Peng}}, \bibinfo {author}
  {\bibfnamefont {B.}~\bibnamefont {Moritz}}, \bibinfo {author} {\bibfnamefont
  {K.}~\bibnamefont {Kummer}}, \bibinfo {author} {\bibfnamefont {N.~B.}\
  \bibnamefont {Brookes}}, \bibinfo {author} {\bibfnamefont {W.~C.}\
  \bibnamefont {Lee}}, \bibinfo {author} {\bibfnamefont {M.}~\bibnamefont
  {Hashimoto}}, \bibinfo {author} {\bibfnamefont {T.}~\bibnamefont {Sarkar}},
  \bibinfo {author} {\bibfnamefont {J.-F.}\ \bibnamefont {He}}, \bibinfo
  {author} {\bibfnamefont {C.~R.}\ \bibnamefont {Rotundu}}, \bibinfo {author}
  {\bibfnamefont {Y.~S.}\ \bibnamefont {Lee}}, \bibinfo {author} {\bibfnamefont
  {R.~L.}\ \bibnamefont {Greene}}, \bibinfo {author} {\bibfnamefont
  {L.}~\bibnamefont {Braicovich}}, \bibinfo {author} {\bibfnamefont
  {G.}~\bibnamefont {Ghiringhelli}}, \bibinfo {author} {\bibfnamefont {Z.~X.}\
  \bibnamefont {Shen}}, \bibinfo {author} {\bibfnamefont {T.~P.}\ \bibnamefont
  {Devereaux}},\ and\ \bibinfo {author} {\bibfnamefont {W.~S.}\ \bibnamefont
  {Lee}},\ }\href {https://doi.org/10.1038/s41586-018-0648-3} {\bibfield
  {journal} {\bibinfo  {journal} {Nature}\ }\textbf {\bibinfo {volume} {563}},\
  \bibinfo {pages} {374} (\bibinfo {year} {2018})}\BibitemShut {NoStop}%
\bibitem [{\citenamefont {Lin}\ \emph {et~al.}(2020)\citenamefont {Lin},
  \citenamefont {Yuan}, \citenamefont {Jin}, \citenamefont {Yin}, \citenamefont
  {Li}, \citenamefont {Zhou}, \citenamefont {Lu}, \citenamefont {Dantz},
  \citenamefont {Schmitt}, \citenamefont {Ding}, \citenamefont {Guo},
  \citenamefont {Dean},\ and\ \citenamefont {Liu}}]{Lin2020}%
  \BibitemOpen
  \bibfield  {author} {\bibinfo {author} {\bibfnamefont {J.}~\bibnamefont
  {Lin}}, \bibinfo {author} {\bibfnamefont {J.}~\bibnamefont {Yuan}}, \bibinfo
  {author} {\bibfnamefont {K.}~\bibnamefont {Jin}}, \bibinfo {author}
  {\bibfnamefont {Z.}~\bibnamefont {Yin}}, \bibinfo {author} {\bibfnamefont
  {G.}~\bibnamefont {Li}}, \bibinfo {author} {\bibfnamefont {K.-J.}\
  \bibnamefont {Zhou}}, \bibinfo {author} {\bibfnamefont {X.}~\bibnamefont
  {Lu}}, \bibinfo {author} {\bibfnamefont {M.}~\bibnamefont {Dantz}}, \bibinfo
  {author} {\bibfnamefont {T.}~\bibnamefont {Schmitt}}, \bibinfo {author}
  {\bibfnamefont {H.}~\bibnamefont {Ding}}, \bibinfo {author} {\bibfnamefont
  {H.}~\bibnamefont {Guo}}, \bibinfo {author} {\bibfnamefont {M.~P.~M.}\
  \bibnamefont {Dean}},\ and\ \bibinfo {author} {\bibfnamefont
  {X.}~\bibnamefont {Liu}},\ }\href {https://doi.org/10.1038/s41535-019-0205-9}
  {\bibfield  {journal} {\bibinfo  {journal} {npj Quantum Materials}\ }\textbf
  {\bibinfo {volume} {5}},\ \bibinfo {pages} {4} (\bibinfo {year}
  {2020})}\BibitemShut {NoStop}%
\bibitem [{\citenamefont {Nag}\ \emph {et~al.}(2020)\citenamefont {Nag},
  \citenamefont {Zhu}, \citenamefont {Bejas}, \citenamefont {Li}, \citenamefont
  {Robarts}, \citenamefont {Yamase}, \citenamefont {Petsch}, \citenamefont
  {Song}, \citenamefont {Eisaki}, \citenamefont {Walters}, \citenamefont
  {García-Fernández}, \citenamefont {Greco}, \citenamefont {Hayden},\ and\
  \citenamefont {Zhou}}]{Nag2020}%
  \BibitemOpen
  \bibfield  {author} {\bibinfo {author} {\bibfnamefont {A.}~\bibnamefont
  {Nag}}, \bibinfo {author} {\bibfnamefont {M.}~\bibnamefont {Zhu}}, \bibinfo
  {author} {\bibfnamefont {M.}~\bibnamefont {Bejas}}, \bibinfo {author}
  {\bibfnamefont {J.}~\bibnamefont {Li}}, \bibinfo {author} {\bibfnamefont
  {H.~C.}\ \bibnamefont {Robarts}}, \bibinfo {author} {\bibfnamefont
  {H.}~\bibnamefont {Yamase}}, \bibinfo {author} {\bibfnamefont {A.~N.}\
  \bibnamefont {Petsch}}, \bibinfo {author} {\bibfnamefont {D.}~\bibnamefont
  {Song}}, \bibinfo {author} {\bibfnamefont {H.}~\bibnamefont {Eisaki}},
  \bibinfo {author} {\bibfnamefont {A.~C.}\ \bibnamefont {Walters}}, \bibinfo
  {author} {\bibfnamefont {M.}~\bibnamefont {García-Fernández}}, \bibinfo
  {author} {\bibfnamefont {A.}~\bibnamefont {Greco}}, \bibinfo {author}
  {\bibfnamefont {S.~M.}\ \bibnamefont {Hayden}},\ and\ \bibinfo {author}
  {\bibfnamefont {K.-J.}\ \bibnamefont {Zhou}},\ }\href
  {https://link.aps.org/doi/10.1103/PhysRevLett.125.257002} {\bibfield
  {journal} {\bibinfo  {journal} {PRL}\ }\textbf {\bibinfo {volume} {125}},\
  \bibinfo {pages} {257002} (\bibinfo {year} {2020})}\BibitemShut {NoStop}%
\bibitem [{\citenamefont {Singh}\ \emph {et~al.}(2022)\citenamefont {Singh},
  \citenamefont {Huang}, \citenamefont {Lane}, \citenamefont {Li},
  \citenamefont {Okamoto}, \citenamefont {Komiya}, \citenamefont {Markiewicz},
  \citenamefont {Bansil}, \citenamefont {Lee}, \citenamefont {Fujimori},
  \citenamefont {Chen},\ and\ \citenamefont {Huang}}]{Singh2022}%
  \BibitemOpen
  \bibfield  {author} {\bibinfo {author} {\bibfnamefont {A.}~\bibnamefont
  {Singh}}, \bibinfo {author} {\bibfnamefont {H.~Y.}\ \bibnamefont {Huang}},
  \bibinfo {author} {\bibfnamefont {C.}~\bibnamefont {Lane}}, \bibinfo {author}
  {\bibfnamefont {J.~H.}\ \bibnamefont {Li}}, \bibinfo {author} {\bibfnamefont
  {J.}~\bibnamefont {Okamoto}}, \bibinfo {author} {\bibfnamefont
  {S.}~\bibnamefont {Komiya}}, \bibinfo {author} {\bibfnamefont {R.~S.}\
  \bibnamefont {Markiewicz}}, \bibinfo {author} {\bibfnamefont
  {A.}~\bibnamefont {Bansil}}, \bibinfo {author} {\bibfnamefont {T.~K.}\
  \bibnamefont {Lee}}, \bibinfo {author} {\bibfnamefont {A.}~\bibnamefont
  {Fujimori}}, \bibinfo {author} {\bibfnamefont {C.~T.}\ \bibnamefont {Chen}},\
  and\ \bibinfo {author} {\bibfnamefont {D.~J.}\ \bibnamefont {Huang}},\ }\href
  {https://link.aps.org/doi/10.1103/PhysRevB.105.235105} {\bibfield  {journal}
  {\bibinfo  {journal} {PRB}\ }\textbf {\bibinfo {volume} {105}},\ \bibinfo
  {pages} {235105} (\bibinfo {year} {2022})}\BibitemShut {NoStop}%
\bibitem [{\citenamefont {Sturm}\ and\ \citenamefont
  {Oliveira}(1989)}]{Sturm1989}%
  \BibitemOpen
  \bibfield  {author} {\bibinfo {author} {\bibfnamefont {K.}~\bibnamefont
  {Sturm}}\ and\ \bibinfo {author} {\bibfnamefont {L.~E.}\ \bibnamefont
  {Oliveira}},\ }\href {https://link.aps.org/doi/10.1103/PhysRevB.40.3672}
  {\bibfield  {journal} {\bibinfo  {journal} {PRB}\ }\textbf {\bibinfo {volume}
  {40}},\ \bibinfo {pages} {3672} (\bibinfo {year} {1989})}\BibitemShut
  {NoStop}%
\bibitem [{\citenamefont {Gibbons}\ and\ \citenamefont
  {Schnatterly}(1977)}]{Gibbons1977}%
  \BibitemOpen
  \bibfield  {author} {\bibinfo {author} {\bibfnamefont {P.~C.}\ \bibnamefont
  {Gibbons}}\ and\ \bibinfo {author} {\bibfnamefont {S.~E.}\ \bibnamefont
  {Schnatterly}},\ }\href {https://link.aps.org/doi/10.1103/PhysRevB.15.2420}
  {\bibfield  {journal} {\bibinfo  {journal} {PRB}\ }\textbf {\bibinfo {volume}
  {15}},\ \bibinfo {pages} {2420} (\bibinfo {year} {1977})}\BibitemShut
  {NoStop}%
\bibitem [{\citenamefont {vom Felde}\ \emph
  {et~al.}(1989{\natexlab{b}})\citenamefont {vom Felde}, \citenamefont
  {Sprösser-Prou},\ and\ \citenamefont {Fink}}]{vom_Felde_1989}%
  \BibitemOpen
  \bibfield  {author} {\bibinfo {author} {\bibfnamefont {A.}~\bibnamefont {vom
  Felde}}, \bibinfo {author} {\bibfnamefont {J.}~\bibnamefont
  {Sprösser-Prou}},\ and\ \bibinfo {author} {\bibfnamefont {J.}~\bibnamefont
  {Fink}},\ }\href {https://doi.org/10.1103/physrevb.40.10181} {\bibfield
  {journal} {\bibinfo  {journal} {Physical Review B}\ }\textbf {\bibinfo
  {volume} {40}},\ \bibinfo {pages} {10181} (\bibinfo {year}
  {1989}{\natexlab{b}})}\BibitemShut {NoStop}%
\bibitem [{\citenamefont {Sing}\ \emph {et~al.}(1999)\citenamefont {Sing},
  \citenamefont {Grigoryan}, \citenamefont {Paasch}, \citenamefont {Knupfer},
  \citenamefont {Fink}, \citenamefont {Lommel},\ and\ \citenamefont
  {Aßmus}}]{Sing1999}%
  \BibitemOpen
  \bibfield  {author} {\bibinfo {author} {\bibfnamefont {M.}~\bibnamefont
  {Sing}}, \bibinfo {author} {\bibfnamefont {V.~G.}\ \bibnamefont {Grigoryan}},
  \bibinfo {author} {\bibfnamefont {G.}~\bibnamefont {Paasch}}, \bibinfo
  {author} {\bibfnamefont {M.}~\bibnamefont {Knupfer}}, \bibinfo {author}
  {\bibfnamefont {J.}~\bibnamefont {Fink}}, \bibinfo {author} {\bibfnamefont
  {B.}~\bibnamefont {Lommel}},\ and\ \bibinfo {author} {\bibfnamefont
  {W.}~\bibnamefont {Aßmus}},\ }\href
  {https://link.aps.org/doi/10.1103/PhysRevB.59.5414} {\bibfield  {journal}
  {\bibinfo  {journal} {PRB}\ }\textbf {\bibinfo {volume} {59}},\ \bibinfo
  {pages} {5414} (\bibinfo {year} {1999})}\BibitemShut {NoStop}%
\bibitem [{\citenamefont {Deng}\ \emph {et~al.}(2013)\citenamefont {Deng},
  \citenamefont {Mravlje}, \citenamefont {Žitko}, \citenamefont {Ferrero},
  \citenamefont {Kotliar},\ and\ \citenamefont {Georges}}]{Deng2013}%
  \BibitemOpen
  \bibfield  {author} {\bibinfo {author} {\bibfnamefont {X.}~\bibnamefont
  {Deng}}, \bibinfo {author} {\bibfnamefont {J.}~\bibnamefont {Mravlje}},
  \bibinfo {author} {\bibfnamefont {R.}~\bibnamefont {Žitko}}, \bibinfo
  {author} {\bibfnamefont {M.}~\bibnamefont {Ferrero}}, \bibinfo {author}
  {\bibfnamefont {G.}~\bibnamefont {Kotliar}},\ and\ \bibinfo {author}
  {\bibfnamefont {A.}~\bibnamefont {Georges}},\ }\href
  {https://link.aps.org/doi/10.1103/PhysRevLett.110.086401} {\bibfield
  {journal} {\bibinfo  {journal} {PRL}\ }\textbf {\bibinfo {volume} {110}},\
  \bibinfo {pages} {086401} (\bibinfo {year} {2013})}\BibitemShut {NoStop}%
\end{thebibliography}%

\end{document}